\documentclass[aip,jcp,preprint]{revtex4-1}

\usepackage[version=3]{mhchem} 
\usepackage{amsmath} 
\usepackage{mathrsfs}
\usepackage{amssymb} 
\usepackage{amsfonts} 
\usepackage{bbold} 
\usepackage{dsfont} 
\usepackage{stmaryrd} 
\usepackage{graphicx} 
\graphicspath{ {./images/} } 
\usepackage{xcolor} 
\usepackage{subcaption}
\usepackage{enumitem} 
\usepackage[colorlinks=true]{hyperref} 
\hypersetup{
    filecolor=black,
    linkcolor=black,
    citecolor=black, 
    urlcolor=black
}

\begin{document}


\title{Approximating committor functions: \\ Objective functions and enhanced sampling}



\author{Thomas Pigeon}
    \email{thomas.pigeon@ifpen.fr}
    \affiliation{IFP Energies Nouvelles, Rond-Point de l’Echangeur de Solaize, BP 3, 69360 Solaize, France}

\author{Gabriel Stoltz}
    \email{gabriel.stoltz@enpc.fr}
    \affiliation{CERMICS, École nationale des ponts et chaussées, Institut Polytechnique de Paris, 6-8 Avenue Blaise Pascal, 77455, Marne-la-Vallée, France}
    \altaffiliation[Also at ]{MATHERIALS team-project, Inria Paris, 48 Rue Barrault, 75013 Paris, France}

\author{Tony Lelièvre}
    \email{tony.lelievre@enpc.fr}
    \affiliation{CERMICS, École nationale des ponts et chaussées, Institut Polytechnique de Paris, 6-8 Avenue Blaise Pascal, 77455, Marne-la-Vallée, France}
    \altaffiliation[Also at ]{MATHERIALS team-project, Inria Paris, 48 Rue Barrault, 75013 Paris, France}


\date{\today}

\begin{abstract}
Many molecular dynamics simulations aim at studying transitions between two states (from reactants to products). In this context, the committor function (which gives for a given molecular configuration the probability to reach the product state before the reactant state) is a pivotal quantity, in particular because it is the optimal importance function for rare event simulation methods such as importance sampling or splitting techniques. These methods are used to sample the reactive path ensemble, and estimate for example the transition rate. However, learning such a function is generally a challenging task due to the high dimensionality of the configuration space. In this work, after reviewing the existing methodologies to construct approximate committor functions, a new loss function based on the application of It\={o}'s formula is proposed to learn the committor function with a minimization procedure on the parameters of a neural network. After comparing this novel approach to existing procedures on the Müller--Brown potential, we introduce a coupling strategy with the Adaptive Multilevel Splitting method to better approximate the committor function using a better sampling of the reactive trajectories. This methodology in which the committor function is iteratively learned only requires initially the knowledge of the reactant and product states.\end{abstract}


\maketitle 

\section*{Introduction}

Rare events are ubiquitous when studying chemical or biological kinetics by molecular dynamics simulation.\cite{PETERS2017} Indeed, the discretization time steps of the simulations are several orders of magnitude smaller than the time scale on which relevant events (such as chemical reactions, protein structural rearrangements, ligand binding or unbinding) occur. As a consequence, the study of such events cannot be achieved using direct molecular dynamics simulations, and dedicated numerical methods have to be devised. 

It is always possible to decompose chemical or biological processes as a succession of reactive events, each of them going from a reactant state $R$ to a product state $P$, with $R \cap P = \emptyset$. Identifying the frequency at which the event $R \rightarrow P$ occurs is the main goal of many methodologies commonly used in molecular simulation. For the study of chemical reactions, the most frequently used methods are based on transition state theory (TST)\cite{Eyring1935, PETERS2017} coupled with a harmonic approximation of the potential energy surface. These methods have the advantage of being computationally cheap; but the harmonic approximation fails to capture many effects in particular for not sufficiently small temperature, or in cases with multiple reactive paths.\cite{Collinge2020} TST can also be used without harmonic approximation,\cite{PETERS2017}
but in any case, TST  overestimates the reaction rate due to the fact that it ignores potential recrossings of the transitions state surface. Moreover, in the case of post transition state bifurcations, TST fails to capture the complexity of a system with multiple products and distinguish the different reaction rates associated with the formation of each product. The correction to this approach requires to compute the transmission coefficients which may be complicated in practical situations.\cite{VandenEijnden2005, PETERS2017} 

Alternatively, one can rely on more computationally intensive methods which aim at directly simulating the transition events. The vast majority of them can be classified into two categories: importance sampling approaches\cite{Hartmann_2012, Dupuis2012, Hartmann2018, hartmann2024} and splitting methods.\cite{Cerou2007, vanErp2003, Allen2005, Allen_2009, Lopes2019, Pigeon2023} In both cases, an importance function is needed and it can be shown that the committor function yields the smallest statistical error.\cite{Cerou2019, Lelievre2016, Lu2014}

Let us consider a system composed of $N$ particles, with a positive definite mass matrix $M \in \mathbb{R}^{3N \times 3N}$ and potential energy $V$. The evolution in time of positions $\textbf{q} \in \Omega$ and momenta $\textbf{p}\in \mathbb{R}^{3N}$ of the system is dictated by a stochastic process such as the Langevin dynamics:
\begin{equation}
	\label{langevin_dynamics}
	\left\{ 
	\begin{aligned}
		\mathrm{d}\textbf{q}_t & = M^{-1}\textbf{p}_t \, \mathrm{d}t, \\
		\mathrm{d}\textbf{p}_t & = -\nabla V(\textbf{q}_t) \, \mathrm{d}t - \gamma \textbf{p}_t \,\mathrm{d}t + \sqrt{\frac{2 \gamma}{\beta}} M^{\frac{1}{2}}\mathrm{d} \, \textbf{B}_t,
	\end{aligned} \right.
\end{equation}
where the friction coefficient $\gamma > 0$ has the dimension of an inverse time, while $\textbf{B}_t$ is a standard $3N$-dimensional Brownian motion. The generator $\mathcal{L}_\mathrm{und}$ of the underdamped Langevin dynamics reads: 
\begin{equation}
	\label{underdamped_generator}
 	\mathcal{L}_{\mathrm{und}} = - \nabla_q V(\textbf{q}) \cdot \nabla_p +  M^{-1} \textbf{p} \cdot \nabla_q - \gamma \textbf{p} \cdot \nabla_p +  \frac{\gamma}{\beta} M : \nabla_p^2,
\end{equation}
where $M:\nabla_p^2 =\displaystyle \sum_{i,j=1}^{3N} M_{ij} \partial^2_{p_i,p_j}$. It can be proven\cite{Rousset2010} that the time rescaled process $(q_{\lambda t})_{t \ge 0}$ converges when $\gamma \to \infty$ to the overdamped Langevin dynamics:
\begin{equation}
	\label{Overdamped_lagevin_dynamics}
	\mathrm{d}\textbf{q}_t = - \nabla_q V(\textbf{q}_t) \, \mathrm{d}t + \sqrt{\frac{2}{\beta}} \, \mathrm{d} \textbf{W}_t,
\end{equation}
where $\textbf{W}_t$ is a standard $3N$ dimensional Brownian motion. The generator $\mathcal{L}_\mathrm{ovd}$ of the overdamped Langevin dynamics reads:
\begin{equation}
	\label{Overdamped_generator}
	\mathcal{L}_{\mathrm{ovd}} = - \nabla_q V \cdot \nabla_q + \frac{1}{\beta} \Delta_q.
\end{equation}
The overdamped Langevin dynamics can also be derived from the underdamped Langevin dynamics~\eqref{langevin_dynamics} by assuming that the mass matrix is proportional to the identity $M=m\mathrm{Id}$, and considering the zero mass limit $m \to 0$.  Even though overdamped dynamics are not adequate to model the dynamics of most systems of interest in chemistry or biology, the committor function associated with these dynamics is often considered as a good collective variable (depending on the positions only), even for whose evolution in time is dictated by the underdamped Langevin dynamics~\eqref{langevin_dynamics}; thus, methods to identify the committor function associated with such dynamics very useful which is why in this work, we will focus on the overdamped dynamics~\eqref{Overdamped_lagevin_dynamics}. 

For such a stochastic dynamics on the positions only, the committor is defined as:
\begin{equation}
    \label{eq:committor_ref}
    \chi(\textbf{q}) = \mathbb{P}_\textbf{q}\left[ \tau_P < \tau_R \right],
\end{equation}
where $\textbf{q}$ is the position in the configuration space, $\tau_R$ (resp. $\tau_P$) is the first time at which the stochastic dynamics governing the time evolution of the system enters the state $R$ (resp.~$P$) and $\mathbb{P}^\textbf{q}$ and means that an average is taken over all realizations of the dynamics starting from fixed positions $\textbf{q}$. This function was introduced in various works such as Refs. \onlinecite{Hummer2004, Vanden-Eijnden2006}, where the extension to the underdamped Langevin dynamics is discussed. The optimality property of the committor function explains why multiple methods (some of them being reviewed in the first section of this paper) were proposed to approximate it.

The committor function $\chi$ defined in~\eqref{eq:committor_ref}, associated with a reactant state $R$ and a product state $P$ for a Markov dynamics with generator $\mathcal L$, satisfies $\mathcal L \chi =0$ in $\left(\overline{R} \cup \overline{P}\right)^\mathrm{c}$, together with Dirichlet boundary conditions on $\overline{R}$ and $\overline{P}$. For the overdamped Langevin dynamics, the partial differential equation (PDE) solved by the committor writes writes:\cite{Vanden-Eijnden2006}
\begin{equation}
\label{back_kolmo_committor}
\left\{
\begin{aligned}
    &\mathcal{L}_{\mathrm{ovd}} \chi(\textbf{q}) = 0, \qquad \forall \,\textbf{q} \in \left(\overline{R} \cup \overline{P}\right)^\mathrm{c} ,\\
    & \chi (\textbf{q}) = 0, \quad \forall \, \textbf{q} \in \overline{R} \; ; \qquad \chi (\textbf{q}) = 1, \quad \forall \, \textbf{q} \in \overline{P}.
\end{aligned}
\right.
\end{equation}
 
As solving the PDE~\eqref{back_kolmo_committor} satisfied by the committor is out of reach for classical methods such as finite element methods due to the high dimensionality of the configuration space, various approaches have been proposed in the literature to build approximate committor functions. In particular, neural networks are currently explored for such approximations in view of the success of such models to learn high dimensional functions. The parameters of a neural network are generally optimized to minimize a loss function. Therefore, to apply such methods, one needs to identify a minimization problem solved by the committor, and to devise a numerical approach to solve this minimization problem. One objective of this work is to review the main methods which have been proposed so far to approximate the committor function using neural network parameterizations. More precisely, we will review three families of methods: 
\begin{enumerate}
    \item[(i)] those relying on~\eqref{eq:committor_ref}, regression techniques and rare event probability estimators;
    \item[(ii)] those relying on Monte Carlo approximations of losses either based on residuals or variational formulations of~\eqref{back_kolmo_committor};
    \item[(iii)] those relying on the fact that, thanks to (5), for any $t>0$, $\exp(t \mathcal{L}_\mathrm{ovd}) \chi = \chi$, and Monte Carlo, approximations of the transition operator $\exp(t \mathcal{L}_\mathrm{ovd})$
\end{enumerate}
All these methods rely on sampling methods, and can thus be included in iterative schemes, which alternate two steps: 
\begin{enumerate}
    \item thanks to estimates of the committor, new samples are obtained using the committor as an importance function in an enhanced sampling method, and
    \item these new samples are in turn used to improve the estimate of the committor.
\end{enumerate}

Besides, a second objective of this work is to introduce and illustrate numerically two new numerical approaches to estimate the committor using NN:
\begin{itemize}
    \item We will construct a new loss function minimised by the committor, based on It\={o}'s formula and numerically study its performance in a simple case where a good sampling of the configuration space is available.
    \item We will study numerically the performance of an iterative learning of the committor function by coupling the learning procedure with the Adaptive Multilevel Splitting (AMS) algorithm\cite{Cerou2007, Lopes2019} to enhance the sampling of reactive trajectories, and thus the quality of the approximation of the committor along reactive paths. This coupling strategy only requires initially the knowledge of the reactant and product states~$R$ and~$P$.  
\end{itemize}

The paper is organized as follows. In Section~\ref{sec:litt_review}, various minimization problems that can be found in the literature to approximate the committor function are presented with a discussion on how to numerically solve them. Then, a novel minimization problem based on an application of It\={o}'s formula is presented in Section~\ref{sec:ito_loss} as well as an iterative learning strategy based on a coupling of this method with the AMS algorithm. Finally, in Section~\ref{sec:results_MB}, some results using the new loss function are presented for the Müller--Brown potential and compared to those obtained by existing methods. Then, in Section~\ref{sec:results_ZP}, the iterative procedure using the minimization of this new loss function and the enhanced sampling method of the reactive paths through AMS is numerically studied on the so-called Z-potential (already considered in Ref.~\onlinecite{Frassek2021}): this potential exhibits a highly non trivial reactive pathway. 

Notice that the objective of this work is not to attack high-dimensional problems: our aim is to illustrate on simple two-dimensional examples the performance of existing techniques, and to introduce new strategies. This should thus be seen as a methodological contribution. In particular, applications of our numerical procedures to the study of catalytic reactions will be the subjects of forthcoming works.

\section{Review of existing methods}
\label{sec:litt_review}

This section aims at presenting the most common methodologies used in the literature to approximate committor functions. First, approaches based on the definition~\eqref{eq:committor_ref} and regression methods are presented. Then, methods based on the fact that the committor function is the solution of the partial derivative equation (PDE)~\eqref{back_kolmo_committor} are discussed in the two following sections. Finally, the last two sections focus on the fact that the committor function is also the solution of a problem involving the transition operator. A non exhaustive list of published works classified by  types of learning objectives to approximate the committor function is given in Table~\ref{table_of_methods}.  

\begin{table}[!ht]
     \centering
     \caption{Classification of some published works using various learning methods to approximate committor functions with neural networks alongside with the type of data required for the training. $\vartheta$ is an arbitrary time-lag, $\tau = \min\left(\tau_R, \tau_P \right)$ is the first hitting time of either $R$ or $P$, $\mu$ is an arbitrary distribution with full support on $\left(\overline{R} \cup \overline{P}\right)^c$ and $\mathrm{e}^{-\beta V} / Z$ is the equilibrium measure of the overdamped process~\eqref{Overdamped_lagevin_dynamics}.}
	\label{table_of_methods}
     \begin{tabular}{p{5cm}|p{5cm}|p{5cm}}
     &  Residuals & Variational\\
      \hline
      Probabilistic definition \newline ($\mathbb{P}_\textbf{q}\left[ \tau_P < \tau_R \right]$) & $\left\{\textbf{q}_0^{i}, \textbf{q}_\tau^{i,1}, .. ,\textbf{q}_\tau^{i,n}\right\}$ \newline with $\left\{\textbf{q}_{0}^i\right\}\sim \mu$ \newline Refs.\onlinecite{Lopes2019, Frassek2021, Jung2023} & N / A \\  
      \hline
      Infinitesimal generator \newline ($\mathcal{L} $) & $\left\{\textbf{q}^i\right\} \sim \mu$ \newline Ref.\onlinecite{yuan2023} & $\left\{\textbf{q}^i\right\} \sim \mathrm{e}^{-\beta V} / Z$ \newline Refs.\onlinecite{Khoo2018, Li2019, rotskoff2020learning, pmlr-v145-rotskoff22a, yuan2023, Kang2024, wang2025}\\
      \hline
      Transition operator \newline ($\mathcal{P}_\vartheta$) & $\left\{\textbf{q}_{0}^i, \textbf{q}_{\vartheta}^{i,1}, .. ,\textbf{q}_{\vartheta}^{i, n}\right\}$ \newline with $\textbf{q}_{0}^i\sim \mu$ \newline Refs.\onlinecite{STRAHAN2023, Strahan_2023, Mitchell2024} & $\left\{\textbf{q}_{0}^i, \textbf{q}_{\vartheta}^i\right\}$ \newline with $\left\{\textbf{q}_{0}^i \right\}\sim \mathrm{e}^{-\beta V} / Z$ \newline Refs.\onlinecite{Li2021, Roux2021, Roux2022, He2022, Chen2023} \\
     \end{tabular}
 \end{table}

 For any learning strategy, some configurations in $\Omega$ are required. Due to the high dimensionality of $\textbf{q}$, it is not possible to create a uniform grid on $\Omega$. Moreover, variational methods require the configurations to be distributed according to the Boltzmann--Gibbs distribution (see Sections~\ref{sec:variationnal_pde}~and~\ref{sec:variational_propagator}). On the other hand, it is not necessary to build a precise approximation of the committor function on the whole of $\Omega$. Indeed, in practice, the reactive trajectories ($R \to P$ or $P \to R$) visit only a restricted portion of $\Omega$, thus in the following by "relevant parts of the configuration space", we refer to the small subset of $\Omega$ visited by the vast majority of these reactive trajectories and the most likely configurations in the vicinity of $R$ and $P$ in the sense of the Boltzmann--Gibbs distribution.  

\subsection{Pointwise approximations and regression}
\label{sec:pointwise_sect}

The committor function can be approximated at various points of the phase space relying on Equation~\eqref{eq:committor_ref} and by directly estimating the probability of reaching $P$ before $R$. Then, a regression task can be performed using the mean absolute error (or the mean squared error) to fit the parameters of a neural network approximating this function at these points.

The most naive approximation approach requires to run multiple trajectories starting from each point. However, those naive Monte Carlo estimators have a high relative variance when the probability to estimate tends to zero (or to one). Indeed, this naive estimator at point $\textbf{q}$ writes:
\begin{equation}
\label{eq:montecarlo_estimator}
    \widehat{\chi}(\textbf{q}) = \frac{n_{\mathrm{success}}}{n},
\end{equation}
where $n$ is the number of launched trajectories starting from the configuration $\textbf{q}$, or number of trials, and $n_\mathrm{success}$ is the number of trajectories reaching $P$ before $R$. As each trial follows a Bernoulli distribution of parameter ${\chi}(\textbf{q})$, the variance of this estimator writes:
\begin{equation}
    \mathrm{Var}\left(\widehat{\chi}(\textbf{q})\right) = \frac{\chi(\textbf{q}) \left(1-\chi(\textbf{q})\right)}{n}.
\end{equation}
Now, the relative error $\sqrt{\mathrm{Var}(\widehat{\chi}(\textbf{q}))} / {\chi}(\textbf{q})$ in the limit ${\chi}(\textbf{q}) \to 0$ behaves as: 
\begin{equation}
    \frac{\sqrt{\mathrm{Var}(\widehat{\chi}((\textbf{q}))}}{\chi(\textbf{q}) } = \sqrt{\frac{\chi(\textbf{q}) - \chi(\textbf{q})^2}{n \chi(\textbf{q})^2}} \approx \sqrt{\frac{1}{n \chi(\textbf{q})}} .
\end{equation}
To obtain an accurate estimation of the committor function, the smaller ${\chi}(\textbf{q})$ is, the larger the number of trials $n$ needs to be; in fact the number of trials should be of order $1/ \chi(\textbf{q})$.

This naive approach thus does not work as such and one needs to rely on enhanced sampling techniques in order to get better estimators of the probability. For example, in this context of the estimation of the committor, two techniques have been used in order to sample the reactive paths and thus obtain a good sampling of the transition region: transition path sampling\cite{Frassek2021, Jung2023} and adaptive multilevel splitting.\cite{Brehier2016, Lopes2019} 

Moreover, another difficulty of this naive approach is the high-dimensionality of the learned function. Therefore, many authors are relying on physics-informed parametrization of the estimator of the committor, by looking for approximations which only depend on a few well chosen collective variables (in which case a simple binning procedure can be used to perform the regression,\cite{Lopes2019}) or by relying on machine learning techniques to actually find good collective variables\cite{Frassek2021} and human-interpretable formulas for the committor.\cite{Jung2023}

\subsection{Residual formulations based on the infinitesimal generator of the dynamics}
\label{sec:residuals_pde}

Another way to approximate the committor is to follow the Physics Informed neural networks (PINN)\cite{raissi2017} approach by defining as loss function the residuals of the PDE~\eqref{back_kolmo_committor} on a given set of points, called collocation points in the PINN litterature. Indeed, thanks to automatic differentiation algorithms, it is possible to evaluate the second order derivatives of the neural network that occur in the left hand side of the PDE~\eqref{back_kolmo_committor} on any set of points. The associated minimization problem reads:
\begin{equation}
\label{PDE_residuals}
    \mathrm{inf} \left\{ \int_{\left(\overline{R} \cup \overline{P}\right)^\mathrm{c}} \left|\mathcal{L}_\mathrm{ovd} f(\textbf{q}) \right|^2  \, \mu(\mathrm{d}\textbf{q}) \, \middle| \, \, f(\textbf{q)} = \mathbb{1}_P(\textbf{q}), \, \forall \, \textbf{q} \in \left(\overline{R} \cup \overline{P}\right) \right\}.
\end{equation}
Here and in the following, since we focus on algorithmic aspects, the functional space in which f is looked for is somewhat implicit, and is in practice a set of $C^\infty$ functions parametrized by a neural network. Let us however emphasize that for~\eqref{PDE_residuals} to be well-posed, one needs $f$ to be sufficiently smooth so that 
\begin{itemize}
    \item[(i)]  $\displaystyle \int |\mathcal{L}f|^2 d\mu < \infty$ and 
    \item[(ii)] the trace of $f$ on $\partial R \cup \partial P$ is well defined
\end{itemize}
For example, $f \in H^2(\mu)$ would be an appropriate space. In particular, if $f$ does not admit a trace on $\partial R \cup \partial P$, a trivial solution to~\eqref{PDE_residuals} would be any constant function on $\left(\overline{R} \cup \overline{P}\right)^c$.

There are two approaches to ensure that the trained networks satisfy the boundary conditions.\cite{barschkis2023} A first option is to enforce them through the functional form of the neural network by considering the minimization problem~\eqref{PDE_residuals} over functions $f$ of the form:
\begin{equation}
    \label{committor_parameterised}
    f_\theta(\textbf{q}) =\widetilde{\mathbb{1}}_{\left(\overline{R} \cup \overline{P}\right)^\mathrm{c}}(\textbf{q})  p_{\theta}(\textbf{q}) + \widetilde{\mathbb{1}}_P(\textbf{q}),
\end{equation}
where $p_\theta$ is the (smooth) neural network with sigmoid output so that $p_{\theta}: \Omega \to (0,1)$, $\theta$ is the vector of parameters of the neural network and $\widetilde{\mathbb{1}}_A$ are approximation of the indicator function, with non zero gradients in a small neighborhood of the boundary of the set $A$.\cite{Li2019} The minimization problem finally reads: 
\begin{equation}
    \label{eq:minimization_PDE_residuals_f}
    \underset{\theta \in \mathbb{R}^{n_\mathrm{p}}}{\mathrm{inf}} \left\{ \int_{\left(\overline{R} \cup \overline{P}\right)^\mathrm{c}} \left|\mathcal{L}_\mathrm{ovd} f_\theta(\textbf{q}) \right|^2 \mu(\mathrm{d}\textbf{q})  \right\}.
\end{equation}

Notice that the function $f_{\theta}$ is smooth on $\Omega$, and thus admits a trace on $\partial R \cup \partial P$. In practice, the loss $\int \mathcal{L}_\mathrm{ovd} f_\theta d\mu$ is approximated by empirical averages over samples and the regions where $\nabla \mathbb{1}_{\left(\overline{R} \cup \overline{P}\right)^c}$ and $\nabla \mathbb{1}_{P}$ are non zero should contain sufficiently many samples so that the boundary conditions are “seen” by the approximation process. Otherwise, as already mentioned above, a constant $p_\theta$ is a solution to the minimization problem based on the discretized loss.

A second approach is to add penalization terms in the loss to enforce the boundary conditions, an approach sometimes referred to as soft constraints. This requires to have a set of collocation points on the boundaries. Parameterizing the approximate committor $\Tilde{\chi}$ by a neural network $g_\theta$ composed of $n_\mathrm{p}$ parameters (hence minimizing over the set of parameters $\theta \in \mathbb{R}^{n_\mathrm{p}}$), the optimization problem with soft constraints writes (compare with~\eqref{eq:minimization_PDE_residuals_f}):
\begin{equation}
    \label{eq:minimization_PDE_residuals_g}
    \underset{\theta \in \mathbb{R}^{n_\mathrm{p}}}{\mathrm{inf}} \left\{ \int_{\left(\overline{R} \cup \overline{P}\right)^\mathrm{c}} \left|\mathcal{L}_\mathrm{ovd} g_\theta(\textbf{q}) \right|^2 \mu(\mathrm{d}\textbf{q})  + \alpha \left(
    \int_{\overline{R}}\left(g_\theta(\textbf{q})\right)^2 \, \mathrm{d}\textbf{q} + \int_{\overline{P}} \left(g_\theta(\textbf{q}) - 1\right)^2 \, \mathrm{d}\textbf{q}\right) \right\},
\end{equation}
where $\alpha > 0$ is a hyper-parameter allowing to explicitly modulate the strength of the constraint. The minimization problem~\eqref{PDE_residuals} is recovered, at least formally, in the limit $\alpha \to +\infty$. This second approach was used for instance in Ref.~\onlinecite{yuan2023} for a 1D particle evolving according to the underdamped langevin dynamics. 

An advantage of this approach is that it can easily be extended to underdamped dynamics, allowing thus to estimate a position and velocity dependent committor function, which might be of interest when studying reaction involving the crossing a very high entropic barriers over a short time (relative to $\gamma^{-1}$).  On the other hand, computing second order derivatives for neural networks with high dimensional inputs requires a substantial computational effort in the training stage of the NN, which is why this approach is not often chosen for high dimensional systems and many studies use the variational formulation presented in the next section. 

Let us emphasize that, in contrast to the variational formulations detailed in Section~\ref{sec:variationnal_pde}, the probability measure $\mu$ can be arbitrary here and long as it has full support on $\left(\overline{R} \cup \overline{P}\right)^\mathrm{c}$. In practice, when using discretized approximations of this loss, a useful approximation of the committor can be obtained if the training configurations (or collocation points) have any distribution as far as it covers the "relevant parts of the configurational space" mentioned before. 

\subsection{Variational formulations based on the infinitesimal generator of the dynamics}
\label{sec:variationnal_pde}

In the particular case of the overdamped Langevin process, many studies are based on a variational formulation of~\eqref{back_kolmo_committor}.\cite{Khoo2018, Li2019, rotskoff2020learning, pmlr-v145-rotskoff22a, yuan2023, Kang2024, wang2025} The method relies on the symmetry of $\mathcal{L}_\mathrm{ovd}$ on $L^2\left(\mathrm{e}^{-\beta V}/Z\right)$, namely: for any smooth functions $f$ and $g$ from $\Omega \to \mathbb{R}, \int (\mathcal{L}_\mathrm{ovd} f) g e^{-\beta V} = \int (\mathcal{L}_\mathrm{ovd} g) f e^{-\beta V} = - \int
\nabla f \cdot \nabla g e^{-\beta V}$. In these approaches, an approximation of the committor is obtained by solving the problem:
\begin{equation}
\label{variationnal_problem_squarred_grad}
    \underset{f}{\mathrm{inf}} \left\{ \int_{\left(\overline{R} \cup \overline{P}\right)^\mathrm{c}} \left|\nabla f(\textbf{q}) \right|^2 \mathrm{e}^{-\beta V(\textbf{q})} \, \mathrm{d}\textbf{q} \, \middle| \, f(\textbf{q)} = \mathbb{1}_P(\textbf{q}), \, \forall \, \textbf{q} \in \left(\overline{R} \cup \overline{P}\right) \right\}.
\end{equation}
The proof that minimizer satisfy~\eqref{back_kolmo_committor} is given in~SI.1. To implement the minimization problem~\eqref{variationnal_problem_squarred_grad} in practice, on can use one of the two options already mentioned in Section~\ref{sec:residuals_pde} to enforce the boundary conditions, and again sampling procedures to approximate $\int_{\left(\overline{R} \cup \overline{P}\right)^\mathrm{c}} \left|\nabla f(\textbf{q}) \right|^2 \mathrm{e}^{-\beta V(\textbf{q})} \, \mathrm{d}\textbf{q}$ by empirical averages over samples distributed according to $\mathrm{e}^{-\beta V(\textbf{q})}/Z$.


In contrast to the PINN approach considered in Section~\ref{sec:variationnal_pde}, less derivatives of $f_\theta$ or $g_\theta$ are needed. However, let us emphasize that in contrast to~\eqref{eq:minimization_PDE_residuals_f} and~\eqref{eq:minimization_PDE_residuals_g}, the probability measure used in the loss function in~\eqref{variationnal_problem_squarred_grad} has to be $\exp(-\beta V) / Z$ in order to get a consistent approximation. This leads to a major difficulty in using such an approach numerically, obtaining a good sampling of the stationary measure of the process, with density proportional to $\mathrm{e}^{-\beta V(\textbf{q})}$ requires in most cases dedicated enhanced sampling strategies.  This is why an active learning approach, in which the approximate committor is used to enhance the sampling and then the neural network is retrained on the newly explored configurations, is considered in Refs.~\onlinecite{pmlr-v145-rotskoff22a, Kang2024, wang2025}.  

\subsection{Residual formulations based on the transition operator of the dynamics}
\label{sec:residuals_propagator}

A fourth approach to formulate a minimization problem solved by the committor function relies on the fact that the committor function is a fixed point of the operator propagating the dynamics forward in time.\cite{Li2021, Roux2021} Let us define the first hitting time of $\overline{R} \cup \overline{P}$ as $\tau = \min(\tau_{R}, \tau_{P})$. The time propagation operator of the dynamics stopped when arriving in $R \cup P$ is then defined for any function $f: \left(\overline{R} \cup \overline{P}\right)^\mathrm{c} \rightarrow \mathbb{R}$ and a given time $t > 0$:
\begin{equation}
    \label{time_propagator}
    \left( \mathcal{P}_t f \right) (\textbf{q}_0) = \mathbb{E}_{\textbf{q}_0} \left[f\left( \textbf{q}_{t \wedge \tau} \right) \right], 
\end{equation}
where $t \wedge \tau = \inf\left\{t, \tau\right\}$, $\textbf{q}_0 \in \left(\overline{R} \cup \overline{P}\right)^\mathrm{c}$ and the expectation is taken with respect to the law of the process~\eqref{Overdamped_lagevin_dynamics}. 

The approach is presented here for the overdamped Langevin dynamics but it can be generalized to the underdamped Langevin dynamics. An application of It\={o}'s formula to the committor function gives:
\begin{equation}
    \label{ito_comm}
    \begin{aligned}
         \mathrm{d}\chi(\textbf{q}_t) &=  \nabla \chi(\textbf{q}_t) \cdot \mathrm{d}\textbf{q}_t + \frac{1}{2} \Delta  \chi(\textbf{q}_t) \frac{2}{\beta} \, \mathrm{d}t, \\
         &=  \mathcal{L}_{\mathrm{ovd}} \chi(\textbf{q}_t) \, \mathrm{d}t + \sqrt{\frac{2}{\beta}} \nabla \chi(\textbf{q}_t) \cdot \mathrm{d}\textbf{W}_t. 
    \end{aligned}
\end{equation}
Given that $\mathcal{L}_\mathrm{ovd}\chi(\textbf{q}) = 0$ for $\textbf{q} \in \left(\overline{R} \cup \overline{P}\right)^\mathrm{c}$, the integration of \eqref{ito_comm} until the time $t \wedge \tau$ starting from a deterministic initial configuration $\textbf{q}_0 \in \left(\overline{R} \cup \overline{P}\right)^\mathrm{c}$ leads to:
\begin{equation}
    \label{ito_comm_2}
        \chi(\textbf{q}_{t \wedge \tau}) - \chi(\textbf{q}_0) = \sqrt{\frac{2}{\beta}} \int_0^{t \wedge \tau} \nabla \chi(\textbf{q}_s) \cdot \mathrm{d}\textbf{W}_s.
\end{equation}
Then, taking the expectation with respect to the law of the process, the term in the right hand side of the previous equality vanishes, so that:
\begin{equation}
    \label{fixed_point_committor}
     \forall t > 0, \, \forall \textbf{q} \in \left(\overline{R} \cup \overline{P}\right)^\mathrm{c}, \quad \chi(\textbf{q}) = (\mathcal{P}_t \chi)(\textbf{q}).
\end{equation}
The operator $\mathcal{P}_t $ can be split into two parts:\cite{Li2021} for any function $f$, 
\begin{equation}
    \label{}
    \begin{aligned}
        \left( \mathcal{P}_t f \right) (\textbf{q}_0) = & \; \mathbb{E}^{\textbf{q}_0}\left[ f\left( \textbf{q}_{t \wedge \tau} \right) \right]\\
        = &\; \mathbb{E}^{\textbf{q}_0}\left[ f\left( \textbf{q}_{t } \right) \mathbb{1}_{t < \tau} \right] + \mathbb{E}^{\textbf{q}_0}\left[ f\left( \textbf{q}_{ \tau} \right) \mathbb{1}_{t \geqslant \tau}  \right]\\ 
        = & \left( \mathcal{P}_t^\mathrm{i} f \right) (\textbf{q}_0) + \left( \mathcal{P}_t^\mathrm{b} f \right) (\textbf{q}_0).
    \end{aligned}
\end{equation}
The part $\mathcal{P}_t^\mathrm{i}$ correspond to the evolution of trajectories staying in $\left(\overline{R} \cup \overline{P}\right)^\mathrm{c}$ ("interior part") while $\mathcal{P}_t^\mathrm{b}$ takes into account trajectories that reach either $R$ or $P$ before time t ("boundary part"). In view of the boundary conditions satisfied by the committor, the equality~\eqref{fixed_point_committor} rewrites: 
\begin{equation}
    \label{committor_fixde_point_2}
    \forall t > 0, \forall \textbf{q} \in \left(\overline{R} \cup \overline{P}\right)^\mathrm{c}, \quad \left( \mathrm{Id} - \mathcal{P}_t^\mathrm{i} \right) \chi (\textbf{q}) - \left( \mathcal{P}_t^\mathrm{b} \mathbb{1}_P \right)(\textbf{q}) = 0.
\end{equation}
It can be checked that the only solution $p$ in the space of measurable functions with values in $[0,1]$ to the equation~\eqref{committor_fixde_point_2} is $\chi$ as detailed in~SI.2. 

A similar identity can be obtained in a discrete time setting and considering the transition operator of the original continuous in time process~\eqref{Overdamped_lagevin_dynamics}. For any bounded measurable function $f : \Omega \to \mathbb{R}$, this operator writes:
\begin{equation}
    \label{eq:not_stopped_transition_operator}
    \forall t \in \mathbb{R}_+ \quad \left(\mathcal {Q}_{t} f \right) (\textbf{q}) = \mathbb{E}^{\textbf{q}_0}\left[f\left(\textbf{q}_{t}\right)\right].
\end{equation}
Let us now introduce a fixed time-lag $\vartheta > 0$ so that we can consider the committor function ${\chi}_\vartheta$ associated with the Markov Chain $(\textbf{q}_n^{\vartheta})_{n \ge 0}$, with transition kernel $\mathcal{Q}_{\vartheta}$. In~SI.3, we prove that this committor function is the only solution of the fixed point equation:
\begin{equation}
\label{eq:fp_discrete_comm}
\left \{ 
\begin{aligned}
    f(\textbf{q}) & = \mathcal Q_{\vartheta} f(\textbf{q}) &  \quad  \textbf{q} \in \left(\overline{R} \cup \overline{P}\right)^\mathrm{c}, \\ 
    f(\textbf{q}) & = \mathbb{1}_P(\textbf{q}) & \quad \textbf{q} \in \left( \overline{R} \cup \overline{P} \right). 
\end{aligned}
\right. 
\end{equation} 


A minimization problem can be obtained by considering the residual associated with the identity~\eqref{committor_fixde_point_2} for a fixed positive time-lag $\vartheta >0$:
\begin{equation}
    \label{minimization_naive_fixed_point}
    \underset{f}{\mathrm{inf}} \left\{ \int_{\left(\overline{R} \cup \overline{P}\right)^\mathrm{c}} \left[ \left( \mathrm{Id} - \mathcal{P}_\vartheta^\mathrm{i} \right) f(\textbf{q}) - \left( \mathcal{P}_\vartheta^\mathrm{b} \mathbb{1}_P \right)(\textbf{q}) \right]^2 \mu(\mathrm{d}\textbf{q}) \right\},
\end{equation}
where $\mu$ is an arbitrary probability measure with full support on $\left(\overline{R} \cup \overline{P}\right)^\mathrm{c}$. Notice that contrary to the minimization problems introduced in the two previous sections, the minimization problem~\eqref{minimization_naive_fixed_point} is set on any function $f$ defined on $\left(\overline{R} \cup \overline{P}\right)^\mathrm{c}$ without specifying any boundary condition. Indeed, any solution to~\eqref{minimization_naive_fixed_point} satisfies~\eqref{committor_fixde_point_2}, and we have shown in~SI.2 that the only solution to~\eqref{committor_fixde_point_2} is indeed the committor. The boundary conditions are in some sense included in the the term $\mathcal{P}_\vartheta^\mathrm{b} \mathbb{1}_P$.

This approach can straightforwardly be adapted to the underdamped Langevin dynamics (and actually to any Markov dynamics), whereas the variational formulation introduced in the next subsection is restricted to time-reversible dynamics.

This method was tested on various low dimensional energy landscapes in Ref. \onlinecite{STRAHAN2023}. Similarly to the PINN approach, the major advantage of this method is that the solution to the minimization problem does not depend on the choice of the probability measure $\mu$ as far as it has full support on $\left(\overline{R} \cup \overline{P}\right)^\mathrm{c}$. In practice, this allows to select initial conditions in the "relevant parts of the phase space" as defined before. On the other hand, the propagators $\mathcal{P}_\vartheta^\mathrm{i}$ and $\mathcal{P}_\vartheta^\mathrm{b}$ have to be estimated by running at least two dynamics starting from every initial condition, see Equation~(11) in Ref.~\onlinecite{STRAHAN2023}. To clarify this point, let us consider:
\begin{equation*}
    \varphi(\textbf{q}_0, \textbf{q}_\vartheta) = f(\textbf{q}_0) - f(\textbf{q}_\vartheta)\mathbb{1}_{\vartheta < \tau} - \mathbb{1}_{P}(\textbf{q}_\tau)\mathbb{1}_{\vartheta \geqslant \tau},
\end{equation*}
so that:
\begin{equation*}
    \left( \mathrm{Id} - \mathcal{P}_\vartheta^\mathrm{i} \right) f(\textbf{q}) - \left( \mathcal{P}_\vartheta^\mathrm{b} \mathbb{1}_P \right)(\textbf{q}) = \mathbb{E}\left[\varphi(\textbf{q}_0, \textbf{q}_\vartheta)\middle|\textbf{q}_0 = \textbf{q}\right].
\end{equation*}
This allows to write the minimization problem~\eqref{minimization_naive_fixed_point} as:
\begin{equation*}
    \underset{f}{\mathrm{inf}} \left\{\mathbb{E}_{\mu}\left[\left(\mathbb{E}_{\textbf{q}_0} \left[\varphi(\textbf{q}_0, \textbf{q}_\vartheta)\right] \right)^2\right] \right\},
\end{equation*}
where $\mathbb{E}_\mu$ refers to the fact that the average is taken over $\textbf{q}_0$ distributed according to the measure $\mu$. In practice, one needs to
approximate $\mathbb{E}_{\mu}\left[\left(\mathbb{E}_{\textbf{q}_0} \left[\varphi(\textbf{q}_0, \textbf{q}_t)\right] \right)^2\right] $ using a sampling
procedure. A first natural idea is to draw independently $n$ configuration $\textbf{q}_0^i$ from $\mu$, and for each of them to run independently $\ell$ dynamics $\left(\textbf{q}_s^{i,j} \right)_{0 \leqslant s \leqslant \vartheta}$ until time $\vartheta$. Then
$\mathbb{E}_{\mu}\left[\mathbb{E}^{\textbf{q}_0}\left[\varphi(\textbf{q}_0, \textbf{q}_t)\right]^2 \right]$ is approximated by 
\[\widehat{\Phi}_{n,\ell}=\frac{1}{n}\sum_{i=1}^n\left(\frac{1}{\ell}\sum_{j=1}^\ell \varphi(\textbf{q}_0^i, \textbf{q}_t^{i,j})\right)^2.\] 
For fixed $\ell$, $\hat \Phi_{n,\ell}$ is a convergent estimator. Indeed, using the law of large
numbers, in the limit in $n\to+\infty$, we obtain:
\begin{equation*}
\begin{aligned}
     \underset{n \to +\infty}{\lim}\frac{1}{n}& \sum_{i=1}^n\left(\frac{1}{\ell}\sum_{j=1}^\ell \varphi(\textbf{q}_0^i, \textbf{q}_\vartheta^{i,j})\right)^2 \\ = &\underset{n \to +\infty}{\lim}\frac{1}{n\ell^2}\sum_{i=1}^n\sum_{j=1}^\ell\sum_{k=1}^\ell \varphi(\textbf{q}_0^i, \textbf{q}_\vartheta^{i,j})\varphi(\textbf{q}_0^i, \textbf{q}_\vartheta^{i,k}) \\
     = & \underset{n \to +\infty}{\lim}\frac{1}{n\ell^2}\sum_{i=1}^n\sum_{j=1}^\ell \left(\varphi(\textbf{q}_0^i, \textbf{q}_\vartheta^{i,j })^2 + 2\sum_{k=j+1}^\ell\varphi(\textbf{q}_0^i, \textbf{q}_\vartheta^{i,j})\varphi(\textbf{q}_0^i, \textbf{q}_\vartheta^{i,k}) \right) \\ 
     = & \frac{1}{\ell}\mathbb{E}_\mu\left[\varphi(\textbf{q}_0, \textbf{q}_\vartheta)^2\right] + \left(1 - \frac{1}{\ell}\right) \mathbb{E}_\mu \left[ \mathbb{E}^{\textbf{q}_0}\left[\varphi(\textbf{q}_0, \textbf{q}_\vartheta)\right]^2\right].
\end{aligned}
\end{equation*}
We see that $n \to \infty$-limit of $\hat \Phi_{n,\ell}$ actually yields a biased estimator of $\mathbb{E}_{\mu}\left[\left(\mathbb{E}_{\textbf{q}_0} \left[\varphi(\textbf{q}_0, \textbf{q}_t)\right] \right)^2\right]$ with a bias vanishing when $\ell \to +\infty$. A more efficient approach, the one used in Ref.~\onlinecite{STRAHAN2023} consists in running only two dynamics from the $n$ configurations drawn independently. More precisely, the authors introduce  
$$\widehat{\Psi}_{n,\ell} = \frac{1}{n}\sum_{i=1}^n \varphi(\textbf{q}_0^i, \textbf{q}_\vartheta^{i,1})\varphi(\textbf{q}_0^i, \textbf{q}_\vartheta^{i,2})$$ 
This estimator converges in the large n limit to $\mathbb{E}_{\mu}\left[\left(\mathbb{E}_{\textbf{q}_0} \left[\varphi(\textbf{q}_0, \textbf{q}_\vartheta)\right] \right)^2\right]$, by the law of large numbers, conditional independence and the Markov property:
\[
\begin{aligned}
    \underset{n \to +\infty}{\lim}\frac{1}{n}\sum_{i=1}^n \varphi(\textbf{q}_0^i, \textbf{q}_\vartheta^{i,1})\varphi(\textbf{q}_0^i, \textbf{q}_\vartheta^{i,2}) & = \mathbb{E}_\mu \left[\varphi(\textbf{q}_0, \textbf{q}_\vartheta^{i,1})\varphi(\textbf{q}_0, \textbf{q}_\vartheta^{2})\right] \\ 
    & = \mathbb{E}_\mu \left[ \mathbb{E}\left[\varphi(\textbf{q}_0, \textbf{q}_\vartheta^{1}) \varphi(\textbf{q}_0, \textbf{q}_\vartheta^{2}) \middle| \textbf{q}_0 \right] \right] \\ 
    & = \mathbb{E}_\mu \left[ \mathbb{E} \left[\varphi(\textbf{q}_0, \textbf{q}_\vartheta^{1}) \middle|\textbf{q}_0 \right] \mathbb{E} \left[\varphi(\textbf{q}_0, \textbf{q}_\vartheta^{2}) \middle| \textbf{q}_0 \right] \right] \\
    & = \mathbb{E}_\mu [  ( \mathbb{E} [\varphi(\textbf{q}_0, \textbf{q}_\vartheta^{1}) |\textbf{q}_0] )^2 ] \\ 
    & = \mathbb{E}_\mu [ \left(\mathbb{E}_{\textbf{q}_0} [ \varphi(\textbf{q}_0,\textbf{q}_\vartheta)] \right)^2 ]
\end{aligned}
\]

We also refer to Ref.~\onlinecite{Strahan_2023} where a variant of this estimator is introduced, using  a reformulation of the problem as an inexact power iteration method, which finally leads to an estimator only requiring a single trajectory.

In vast regions of configuration space close to the subsets $R$ (resp. $P$), the values of the committor function are close to $0$ (resp. $1$). This leads to numerical difficulties to converge to accurate approximations of the committor. To address this problem, the authors of Ref.~\onlinecite{Mitchell2024} propose to build a loss function by rewriting equation~\eqref{committor_fixde_point_2} as: 
\begin{equation*}
    \forall t > 0, \forall \textbf{q} \in \left(\overline{R} \cup \overline{P}\right)^\mathrm{c}, \quad \ln \left(\chi (\textbf{q})\right) = \ln \left( \mathcal{P}_t^\mathrm{i}  \chi (\textbf{q})\right)  + \ln \left( \mathcal{P}_t^\mathrm{b} \mathbb{1}_P (\textbf{q})\right). 
\end{equation*}
To alleviate the numerical difficulties in the vicinity of $P$, the same equation can be written for the "backward committor" $\chi_{P \to R} = 1 - \chi$: 
\begin{equation*}
    \forall t > 0, \forall \textbf{q} \in \left(\overline{R} \cup \overline{P}\right)^\mathrm{c}, \quad \ln \left(\chi_{P \to R} (\textbf{q})\right) = \ln \left( \mathcal{P}_t^\mathrm{i}  \chi_{P \to R} (\textbf{q})\right)  + \ln \left( \mathcal{P}_t^\mathrm{b} \mathbb{1}_R (\textbf{q})\right). 
\end{equation*}
The final loss addressing the numerical difficulties in the vicinity of $R$ and $P$ is then the sum of the residual of these two identities. for the sake of clarity, we only write the minimization problem for the "forward" committor:
\begin{equation}
    \label{eq:minimization_fixed_point_log}
    \underset{f}{\mathrm{inf}} \left\{ \int_{\left(\overline{R} \cup \overline{P}\right)^\mathrm{c}} \left[ \ln{\left(f(\textbf{q})\right)} - \ln{\left(\mathcal{P}_\vartheta^\mathrm{i}f(\textbf{q}) + \mathcal{P}_\vartheta^\mathrm{b} \mathbb{1}_P (\textbf{q}) \right)}\right]^2 \mu(\mathrm{d}\textbf{q}) \right\}.
\end{equation}

To build an estimator of the functional to be minimized in~\eqref{eq:minimization_fixed_point_log}, here again, we draw independently $n$ configuration $\textbf{q}_0^i$ from $\mu$, and for each of them we run independently $\ell$ dynamics $(\textbf{q}^{i,j}_\vartheta)_{\vartheta
\ge 0}$ until time~$\vartheta$. Then, considering:
\begin{equation*}
    \widehat{\phi}_\ell(\textbf{q}_0^i) = \ln{\left(\frac{1}{\ell} \sum_{j=1}^\ell \left(f(\textbf{q}_\vartheta^{i,j})\mathbb{1}_{\vartheta < \tau} + \mathbb{1}_{P}(\textbf{q}_\vartheta^{i,j})\mathbb{1}_{\vartheta \geqslant \tau}\right)\right)},
\end{equation*}
we can minimize the estimator:
\begin{equation*}
    \widehat{\Upsilon}_{n,\ell} = \frac{1}{n}\sum_{i=1}^n \left(\ln f(\textbf{q}_0^i) -  \widehat{\phi}_\ell(\textbf{q}_0^i)\right)^2.
\end{equation*}

For a fixed initial condition $\textbf{q}_0^i$,
one has $\underset{{l \to \infty}}{\lim} \hat
\phi_l = \ln ( \mathcal{P}^i_\vartheta f(\textbf{q}_0^i) +
\mathcal{P}^b_\vartheta \mathbb{1}_P (\textbf{q}_0^i))$, but due to Jensen inequality, for a fixed $l$, notice that $\mathbb{E}(\hat \phi_l) < \ln (\mathcal{P}^i_\vartheta f(\textbf{q}_0^i) + \mathcal{P}^b_\vartheta \mathbb{1}_P(\textbf{q}_0^i))$.
Ref.~\onlinecite{Mitchell2024} discusses the impact of this
bias on the results, and demonstrates numerically that the loss including a logarithm yields better results that losses without the log function.


\subsection{Variational formulations based on the transition operator of the dynamics}
\label{sec:variational_propagator}


For overdamped Langevin dynamics, it is possible to write a variational formulation of~\eqref{committor_fixde_point_2} relying on the symmetry of the operator $\mathrm{Id} - \mathcal{P}_t^\mathrm{i}$ in $L^2\left(\mathrm{e}^{-\beta V}/Z\right)$, as done by Li et al. in Ref. \onlinecite{Li2021}. More precisely, the committor solves the following minimization problem:
\begin{equation}
    \label{minimization_fixed_point_canonical}
        \underset{f}{\mathrm{inf}} \left\{ \frac{1}{2} \int_{\left(\overline{R} \cup \overline{P}\right)^\mathrm{c}}  f(\textbf{q})\left(\mathrm{Id} - \mathcal{P}_\vartheta^\mathrm{i} \right) f (\textbf{q}) \mathrm{e}^{-\beta V(\textbf{q})} \, \mathrm{d}\textbf{q}  - \int_{\left(\overline{R} \cup \overline{P}\right)^\mathrm{c}} f(\textbf{q}) \mathcal{P}_\vartheta^\mathrm{b} \mathbb{1}_P(\textbf{q}) \mathrm{e}^{-\beta V(\textbf{q})} \, \mathrm{d}\textbf{q} \right\}.
\end{equation}
It is shown in~SI.4 that a solution to the minimization problem~\eqref{minimization_fixed_point_canonical} satisfies~\eqref{committor_fixde_point_2}, and is therefore the committor function. 

Using the Euler--Maruyama integration scheme   approximation of~\eqref{Overdamped_lagevin_dynamics}, which writes
\begin{equation}
    \label{discretized_overdamped_langevin_euler_maruyama}
    Q^{\ell+1} = Q^{\ell} -\nabla V(Q^\ell) \Delta t + \sqrt{\frac{2 \Delta t}{\beta}} G^{\ell+1},
\end{equation}
where $\left(G^\ell\right)_{1 \leqslant \ell \leqslant L}$ is a sequence of independent Gaussian random variables with zero mean and identity covariance matrix, it is possible to get a consistent (for sufficiently small timestep) discretization of the functional in~\eqref{minimization_fixed_point_canonical} using a single (long) trajectory $\left(Q^\ell\right)_{1 \leqslant \ell \leqslant L}$ thanks to the ergodicity of this dynamics with respect to the Boltzmann--Gibbs measure. Assuming that the time-lag $\vartheta$ is a multiple of the discretization time-step $\vartheta = k\Delta t$, it writes: 
\begin{equation}
\label{eq:discrtized_minimization_fixed_point_canonical}
\frac{1}{L-k} \sum_{\ell = 1}^{L-k} \mathbb{1}_{\left(\overline{R} \cup \overline{P}\right)^\mathrm{c}}(Q^\ell)\left(\mathbb{1}_{\left(\overline{R} \cup \overline{P}\right)^\mathrm{c}}(Q^{\ell +k})\frac{1}{2} f(Q^\ell) \left(f(Q^\ell) - f(Q^{\ell+k}) \right) - \mathbb{1}_{\overline{R} \cup \overline{P}}(Q^{\ell +k})f(Q^\ell)\mathbb{1}_P(Q^{\ell +k})\right).
\end{equation}
The propagators $\mathcal{P}_t^\mathrm{i}$ and $\mathcal{P}_t^\mathrm{b}$ are simply estimated using single trajectory data, which is possible only due to the fact that the integration is performed against the equilibrium measure of the overdamped process. In practice, the function $f$ is approximated by a neural network, and the optimization is performed over the parameters of the NN.

The approach proposed by Roux et al. in Ref.~\onlinecite{Roux2021} and further used in various subsequent works\cite{He2022, Roux2022, Chen2023, chen2025, Megias2025, arredondo2025} is quite similar in terms of required data to the one in Equation~\eqref{minimization_fixed_point_canonical}. Indeed, the authors propose formulate the problem as:
\begin{equation}
 \label{eq:minimization_fixed_point_roux}
    \underset{f}{\mathrm{inf}} \left\{\underset{n \to +\infty}{\lim}\frac{1}{T-\vartheta}\int_{0}^{T - \vartheta} \left( f(\textbf{q}_{t + \vartheta})  - f(\textbf{q}_t)\right)^2 \, \mathrm{d}t  \, \middle| \, f(\textbf{q)} = \mathbb{1}_P(\textbf{q}), \, \forall \, \textbf{q} \in \left(\overline{R} \cup \overline{P}\right)  \right\}.
\end{equation}
Let us emphasize that in ~\eqref{eq:minimization_fixed_point_roux}, and contrary to~\eqref{minimization_fixed_point_canonical}, the minimization is performed over functions which satisfy the boundary conditions. This problem can be rewritten as:
\begin{equation}
 \label{eq:minimization_fixed_point_roux_2}
    \underset{f}{\mathrm{inf}} \left\{\int_{\Omega} \left( \mathcal {Q}_{\vartheta} \left(f^2\right)(\textbf{q})  -2 f(\textbf{q})\mathcal {Q}_{\vartheta} f(\textbf{q}) +f(\textbf{q})^2\right) \mathrm{e}^{-\beta V(\textbf{q})} \, \mathrm{d}\textbf{q}  \, \middle| \, f(\textbf{q)} = \mathbb{1}_P(\textbf{q}), \, \forall \, \textbf{q} \in \left(\overline{R} \cup \overline{P}\right)  \right\}.
\end{equation}
In~SI.5, we provide a proof that a solution of the problem~\eqref{eq:minimization_fixed_point_roux_2} or~\eqref{eq:minimization_fixed_point_roux} is the committor function associated with the discrete Markov-Chain defined as the unique solution to~\eqref{eq:fp_discrete_comm}.

Concerning the numerical implementation of such an approach, one can simply use the parameterization~\eqref{committor_parameterised} enforcing the boundary conditions and use a very long trajectory of a discretized overdamped Langevin dynamics to approximate the functional in~\eqref{eq:minimization_fixed_point_roux}. 

The major difficulty to use a numerical approach based on either the minimization problem~\eqref{minimization_fixed_point_canonical} or~\eqref{eq:minimization_fixed_point_roux_2} is to obtain a trajectory of the overdamped process long enough so that the relevant part of configuration space, in the sense defined before, are visited. One could use biased potentials to enhance the sampling, see for instance Ref.\onlinecite{Lelievre2016}.



\section{A new objective function to learn the committor}
\label{sec:ito_loss}

The aim of this section is to propose an alternative approach to approximate the committor function using neural networks. First, an alternative minimization problem is presented in Section~\ref{sec:ito_loss_continuous} in a continuous-time setting. Section~\ref{sec:ito_loss_discretized} is dedicated to the discretization of the objective function. Finally Section~\ref{sec:ito_loss_with_ams} focuses on an iterative workflow in which, alternatively, AMS simulations are run to sample trajectories and the training of the approximate committor is performed using these trajectories. 

\subsection{Minimization problem}
\label{sec:ito_loss_continuous}

We propose, in this section, a family of minimization problems solved by the committor function, different from the ones considered in Section~\ref{sec:litt_review}.  The approach is presented for the overdamped Langevin process but can be easily extended to underdamped Langevin dynamics. 

We first introduce a strictly monotonous function $h \in C^2(I)$, where $I$ is an interval of $\mathbb{R}$, whose aim is to put emphasis on certain range of values of the committor function. Indeed in large regions around $R$ and $P$, the values of the committor function are close to 0 or 1 respectively. A rescaling of the committor function in order to stretch its variations in these regions is therefore useful in practice. For example, considering the logarithm of the committor helps to get a better description of the committor when it takes values close to zero, and thus leads to better conditioned minimization problem as introduced in Ref.~\onlinecite{Mitchell2024} and discussed in Section~\ref{sec:variational_propagator}. Let us consider two reals numbers $(\mathrm{a}, \mathrm{b})$ with $\left( \min(\mathrm{a}+\mathrm{b}, \mathrm{b}), \max(\mathrm{a}+\mathrm{b}, \mathrm{b})\right) \subset I$ so that for all $p \in [0,1]$,
$a p + b \in I$. Then:
\begin{equation*}
    \begin{aligned}
        \mathcal{L}_{\mathrm{ovd}}&h(\mathrm{a}~\chi(\textbf{q}) + \mathrm{b}) \\ & =   -\nabla V(\textbf{q}) \cdot \nabla \left[h(\mathrm{a}~\chi(\textbf{q}) + \mathrm{b})\right]  + \frac{1}{\beta} \Delta \left[h(\mathrm{a}~\chi(\textbf{q})  + \mathrm{b})\right] \\
        & =  - \mathrm{a}~h'(\mathrm{a}~\chi(\textbf{q}) + \mathrm{b}) \nabla V(\textbf{q}) \cdot \nabla \chi(\textbf{q}) \\ 
        & \qquad \qquad  + \frac{1}{\beta} \nabla \cdot \left[\mathrm{a}~h'(\mathrm{a}~\chi(\textbf{q}) + \mathrm{b}) \nabla \chi(\textbf{q}) \right] \\ 
        & =  \mathrm{a}~h'(\mathrm{a}~\chi(\textbf{q}) + \mathrm{b}) \mathcal{L}_{\mathrm{ovd}}\chi(\textbf{q}) + \frac{\mathrm{a}^2}{\beta} h''\left(\mathrm{a}~\chi(\textbf{q}) + \mathrm{b}\right) \left| \nabla \chi(\textbf{q})\right|^2.
    \end{aligned}
\end{equation*}
Given the partial differential equation~\eqref{back_kolmo_committor} satisfied by $\chi$, we finally obtain:
\begin{equation}
    \label{back_kolmo_adapted}
    \mathcal{L}_{\mathrm{ovd}}h(\mathrm{a}~\chi(\textbf{q}) + \mathrm{b}) = \frac{\mathrm{a}^2}{\beta} h''\left(\mathrm{a}~\chi(\textbf{q}) + \mathrm{b}\right)\left| \nabla \chi(\textbf{q})\right|^2.
\end{equation}
Then, It\={o}'s formula leads to: 
\begin{equation*}
\begin{aligned}
    \mathrm{d}h\left(\mathrm{a}~\chi(\textbf{q}_t) + \mathrm{b}\right) =  & ~~ \frac{\mathrm{a}^2}{\beta} h''\left(\mathrm{a}~\chi(\textbf{q}_t) + \mathrm{b}\right) \left| \nabla \chi(\textbf{q}_t)\right|^2 \mathrm{d}t \\ & +  \mathrm{a}\sqrt{\frac{2}{\beta}}h'(\mathrm{a}~\chi(\textbf{q}_t) + \mathrm{b})\nabla \chi(\textbf{q}_t) \cdot \mathrm{d}\textbf{W}_t.
\end{aligned}
\end{equation*}
By integrating until time $\vartheta \wedge \tau$, for any $\textbf{q}_0 \in \left(\overline{R} \cup \overline{P}\right)^\mathrm{c}$, we therefore obtain: 
\begin{equation*}
    \label{eq:ito_integrated}
    \begin{aligned}
    h(\mathrm{a}~&\chi(\textbf{q}_\vartheta) + \mathrm{b}) \mathbb{1}_{\vartheta < \tau} +  h(\mathrm{a}~\mathbb{1}_P(\textbf{q}_\tau) + \mathrm{b}) \mathbb{1}_{\vartheta \geqslant \tau} - h(\mathrm{a}~\chi(\textbf{q}_0) + \mathrm{b}) \\ = & ~~ \frac{\mathrm{a}^2}{\beta}\int_0^{\vartheta \wedge \tau} h''(\mathrm{a}~\chi(\textbf{q}_s) + \mathrm{b}) \left| \nabla \chi(\textbf{q}_s)\right|^2 \mathrm{d}s \\ & + \mathrm{a}\sqrt{\frac{2}{\beta}} \int_0^{\vartheta \wedge \tau}h'(\mathrm{a}~\chi(\textbf{q}_s) + \mathrm{b})\nabla \chi(\textbf{q}_s) \cdot \mathrm{d}\textbf{W}_s.
    \end{aligned}
\end{equation*}
Penalizing the residual of this identity for any initial conditions $\textbf{q}_0 \in \left(\overline{R} \cup \overline{P}\right)^\mathrm{c}$ distributed according to some probability measure $\mu$, the following expression is obtained:
\begin{equation}
    \label{ito_comm_4}
    \begin{aligned}
        \int_{\left(\overline{R} \cup \overline{P}\right)^\mathrm{c}} \mathbb{E}_{\textbf{q}_0}\Bigg( &h(\mathrm{a}~\chi(\textbf{q}_\vartheta) + \mathrm{b}) \mathbb{1}_{\vartheta < \tau} + h(\mathrm{a}~\mathbb{1}_P(\textbf{q}_\tau) + \mathrm{b}) \mathbb{1}_{\vartheta \geqslant \tau} - h(\mathrm{a}~\chi(\textbf{q}_0) + \mathrm{b})\\ & - \frac{\mathrm{a}^2}{\beta}\int_0^{\vartheta \wedge \tau} h''(\mathrm{a}~\chi(\textbf{q}_s) + \mathrm{b}) \left| \nabla \chi(\textbf{q}_s)\right|^2 \mathrm{d}s \\ & - \mathrm{a}\sqrt{\frac{2}{\beta}} \int_0^{\vartheta \wedge \tau}h'(\mathrm{a}~\chi(\textbf{q}_s) + \mathrm{b})\nabla \chi(\textbf{q}_s) \cdot \mathrm{d}\textbf{W}_s\Bigg)^2 \mu(\mathrm{d}\textbf{q}_0) = 0.
    \end{aligned}
\end{equation}
Here again, we stress that the measure $\mu$ is arbitrary as long as it has a full support on $\left(\overline{R} \cup \overline{P}\right)^\mathrm{c}$. Therefore, the committor function is the solution of the minimization problem:
\begin{equation}
    \label{minimization_ito_lemma}
    \begin{aligned}
     \underset{f}{\mathrm{inf}} \int_{\left(\overline{R} \cup \overline{P}\right)^\mathrm{c}} \mathbb{E}_{\textbf{q}_0} \Bigg(&h(\mathrm{a}~f(\textbf{q}_\vartheta) + \mathrm{b}) \mathbb{1}_{\vartheta < \tau} + h(\mathrm{a}~\mathbb{1}_P(\textbf{q}_\tau) + \mathrm{b}) \mathbb{1}_{t \geqslant \tau} - h(\mathrm{a}~f(\textbf{q}_0) + \mathrm{b)} \\ & - \frac{\mathrm{a}^2}{\beta}\int_0^{\vartheta \wedge \tau} h''(\mathrm{a}~f(\textbf{q}_s) + \mathrm{b}) \left| \nabla f(\textbf{q}_s)\right|^2 \mathrm{d}s \\ & - \mathrm{a}\sqrt{\frac{2}{\beta}} \int_0^{\vartheta \wedge \tau}h'(\mathrm{a}~f(\textbf{q}_s) + \mathrm{b})\nabla f(\textbf{q}_s) \cdot \mathrm{d}\textbf{W}_s \Bigg)^2\mu(\mathrm{d}\textbf{q}_0),   
    \end{aligned}
\end{equation}
where the trajectories satisfy~\eqref{Overdamped_lagevin_dynamics}. Let us emphasize that the unique minimizer of~\eqref{minimization_ito_lemma} among functions $f: \left(R \cup P \right)^\mathrm{c} \to \mathbb{R}$ is the committor function, thanks to the strict monotonicity of $h$, and the fact that the boundary conditions $f=1$ on $\partial P$ and $f=0$ on $\partial R$ are implicitly included in the formulation.

As for all other approaches mentioned in the previous section, a key point for the applicability of these methods is to train the committor on a dataset of configurations covering the relevant parts of the configuration space. This problem is most often solved by introducing an active learning procedure in which the committor function is successively trained and then used to enhance the sampling of configurations. In Sections~\ref{sec:ito_loss_with_ams} and~\ref{sec:results_ZP}, we provide details about the implementation of this iterative learning procedure for the adaptive multilevel splitting technique, and we illustrate numerically the efficiency of this approach for the sampling of rare configurations and reactive trajectories. This relies on the fact that AMS allows to sample unbiased reactive trajectories.\cite{Brehier2016, Lopes2019}

\paragraph{Remark 1.}

\textit{The method presented here can be extended to biased dynamics of the form:}
\begin{equation}
	\label{biased_Overdamped_lagevin_dynamics}
	\mathrm{d}\textbf{q}_t = - \nabla_q (V - U)(\textbf{q}_t)\mathrm{d}t + \sqrt{\frac{2}{\beta}} \mathrm{d}\textbf{W}_t.
\end{equation}
\textit{The bias $U$ can be for instance (a fraction of) the free energy associated with some collective variables. The generator }$\mathcal{L}_U$ \textit{of the dynamics~\eqref{biased_Overdamped_lagevin_dynamics} writes:}
\begin{equation}
	\label{biased_Overdamped_generator}
	\mathcal{L}_U = - \nabla_q (V - U) \cdot \nabla_q + \frac{1}{\beta} \Delta_q = \mathcal{L}_\mathrm{ovd} + \nabla_q U \cdot \nabla_q.
\end{equation}
\textit{The same derivation as the one leading to~\eqref{minimization_ito_lemma} allows to check that the committor function is the solution to the following minimization problem:} 
\begin{equation}
    \label{minimization_ito_lemma_biased}
    \begin{aligned}
     \underset{f}{\mathrm{inf}} \int_{\left(\overline{R} \cup \overline{P}\right)^\mathrm{c}} \Bigg(&h(\mathrm{a}\,f(\textbf{q}_\vartheta) + \mathrm{b}) \mathbb{1}_{\vartheta < \tau} + h(\mathrm{a}\,\mathbb{1}_P(\textbf{q}_\tau) + \mathrm{b}) \mathbb{1}_{\vartheta \geqslant \tau} - h(\mathrm{a}\,f(\textbf{q}_0) + \mathrm{b)} \\ 
     & - \frac{\mathrm{a}^2}{\beta}\int_0^{\vartheta \wedge \tau} h''(\mathrm{a}\,f(\textbf{q}_s) + \mathrm{b}) \left| \nabla f(\textbf{q}_s)\right|^2 \, \mathrm{d}s \\
     & - \mathrm{a} \int_0^{\vartheta \wedge \tau} h'(\mathrm{a}\,f(\textbf{q}_s) + \mathrm{b})\nabla f(\textbf{q}_s) \cdot \nabla U (\textbf{q}_s) \, \mathrm{d}s \\
     & - \mathrm{a}\sqrt{\frac{2}{\beta}} \int_0^{\vartheta \wedge \tau}h'(\mathrm{a}\,f(\textbf{q}_s) + \mathrm{b})\nabla f(\textbf{q}_s) \cdot \mathrm{d}\textbf{W}_s \Bigg)^2\mu(\mathrm{d}\textbf{q}_0),   
    \end{aligned}
\end{equation}
\textit{where the trajectories are generated from the process~\eqref{biased_Overdamped_lagevin_dynamics}. The interest of this approach is that the biasing potential can be used to facilitate the exploration of the configuration space. We have checked (numerical results not reported here) that this approach yields consistent results on the Müller-Brown example from Section~\ref{sec:MB_results}, using a biasing potential $U$ of the form: 
\begin{equation}
\label{eq:U_potential}
U(\textbf{q}) = \frac{V(\textbf{q}) - E_\mathrm{cut}}{1 + \mathrm{e}^{\lambda\left(V(\textbf{q}) - E_\mathrm{cut}\right)}}
\end{equation}
so that $V-U$ is flattened on regions where $V < E_{cut}$.
}

\subsection{Loss function in a discrete time setting}
\label{sec:ito_loss_discretized}

Let us now assume that $K$ trajectories $Q^k = \left(Q^{k,\ell}\right)_{0 \leq \ell \leq L}$ of $L$ time steps of the discretized overdamped Langevin process~\eqref{discretized_overdamped_langevin_euler_maruyama} with initial conditions drawn independently from $\mu$ (see Equation~\eqref{minimization_ito_lemma}) are available. We will use the Euler--Maruyama discretization~\eqref{discretized_overdamped_langevin_euler_maruyama} introduced before. We define $\ell_k$ as the smallest integer such that $Q^{k,\ell_k} \in \overline{R} \cup \overline{P}$ so that $L \wedge \ell_k$ is the length of the $k$-th realization of the dynamics~\eqref{discretized_overdamped_langevin_euler_maruyama} stopped when it reaches $R$ or $P$. This allows to write the discretized loss function approximating~\eqref{minimization_ito_lemma} as:
\begin{equation}
    \label{loss_ito}
    \begin{aligned}
     \mathscr{L}({\theta})
    =  \frac{1}{K}\sum_{k=1}^K \Biggl(&h(\mathrm{a}~f_\theta(Q^{k, L \wedge \ell_k }) + \mathrm{b}) - h(\mathrm{a}~f_\theta(Q^{k,0}) + \mathrm{b}) \\ &  - \frac{\mathrm{a}^2 \Delta t}{\beta} \sum_{\ell=1}^{L \wedge \ell_k } h''(\mathrm{a}~f_\theta(Q^{k,\ell}) + \mathrm{b}) | \nabla f_\theta(Q^{k, \ell}) |^2 \\ & - \mathrm{a} \sqrt{\frac{2 \Delta t}{\beta}}  \sum_{\ell=1}^{L \wedge \ell_k } h'(\mathrm{a}~f_\theta(Q^{k,\ell}) + \mathrm{b})\nabla f_\theta(Q^{k,\ell}) \cdot G^{k,\ell+1}   \Biggl)^2.  
    \end{aligned}
\end{equation}
We will discuss in details the choice of the distribution $\mu$ of the initial conditions $Q_0^k$ in Sections~\ref{sec:ito_loss_with_ams} and~\ref{sec:results_ZP}. The quality of the trained committor function obviously depends on whether this distribution of initial points covers all the relevant parts of the space or not. 

The function $h$, as well as the numbers $\mathrm{a}$ and $\mathrm{b}$, must be chosen appropriately depending on the system studied and the set of trajectories at hand. We explore two options for the choice of function $h$ in the numerical tests. 

\paragraph{Loss function with identity.} The first option is to simply use $h(z) = z$. In that particular case, there is no point in using various values of the numbers $\mathrm{a}$ and $\mathrm{b}$, thus they were fixed to $1$ and $0$ respectively. 

\paragraph{Loss function with logarithm.} The second explored option is to use $h(z) = \ln (z)$ as in Ref.~\onlinecite{Mitchell2024}. This is a natural choice since it allows to put emphasis on low probability regions. To put also some emphasis on regions close to $P$ (where the probability is close to $1$), the actual loss is built as the sum of two losses, the first one built with $h = \ln$, $\mathrm{a} = 1$, $\mathrm{b}=\varepsilon$, and the second one with $h = \ln$, $\mathrm{a} = -1$, $\mathrm{b}=1 + \varepsilon$. We introduced the parameter $\varepsilon> 0$ to prevent numerical issues. In Section~\ref{sec:num_results}, this parameter is fixed given the prior knowledge of the order of magnitude of the smallest/largest probabilities one would like to estimate around the reactant and product states.  


\subsection{Iterative learning scheme}
\label{sec:ito_loss_with_ams}

For given reactant and product states $R$ and $P$, a good approximation of the committor function typically requires samples in the transition regions: these are usually not available. In order to remedy this issue, one can consider an iterative approach to successively learn an approximation of the committor function using a given ensemble of samples, and to then use an enhanced sampling method based on this approximation of the committor function to get samples in new regions. In this section, we present such an approach combining an approximation of the committor function using the loss~\eqref{loss_ito} with the AMS algorithm\cite{Cerou2007, Lopes2019} to enhance the sampling of reactive trajectories. This method will be tested on a toy example in Section~\ref{sec:results_ZP}.

The iterative learning scheme requires to define a first approximation of the committor function. This is obtained by considering data points in the vicinity of the states $R$ and $P$. These data points could typically be the points visited along the trajectories used to sample initial conditions for AMS. One can then minimize the loss~\eqref{loss_ito} to get a first approximation of the committor function, which will be used as an importance function for the first AMS runs.

The iterative procedure then successively applies the AMS algorithm with the current approximation of the committor function to get new samples, and a minimization of the loss~\eqref{loss_ito} to refine the approximation. Let us describe these two steps.

The AMS algorithm detailed in~SI.6 can be summarized as follows: 
\begin{enumerate}
    \item Initialization: sample $N_\mathrm{rep}$ initial conditions from the exit distribution of $R$; run $N_\mathrm{rep}$ trajectories from the initial conditions until either $R$ or $P$ is reached;
    \item Iterations: while some of the $N_\mathrm{rep}$ trajectories finish in $R$ rather than $P$, update the ensemble of $N_\mathrm{rep}$ trajectories by replacing the $k$ worst trajectories in terms of maximal progress along an importance function $\xi$, by $k$ new trajectories obtained by a resampling procedure among the $N_\mathrm{rep}-k$ remaining ones.
\end{enumerate}
The maximal progress of each trajectory is measured using as an importance function the current approximation of the committor function. Indeed, it can be shown that the optimal importance function for the AMS algorithm (in terms of asymptotic variance)\cite{Cerou2019} is the committor function.

In practice, we actually use a couple of AMS runs at each iteration: one "forward" with trajectories starting from $R$ (to sample $R \rightarrow P$ trajectories), and one "backward" with trajectories starting from $P$ (to sample $P \rightarrow R$ trajectories). Recall that, for the overdamped Langevin dynamics, the "forward" ($\chi$) and "backward"(${\chi}_{P \rightarrow R}$) committor functions are related via:
\begin{equation}
    \label{backward_and_forward_relation}
    {\chi}(\textbf{q}) = 1 -{\chi}_{P \rightarrow R}(\textbf{q}),
\end{equation}
so that we only use an approximation of the forward committor to define both importance functions.

Let us now turn to the update of the approximation of the committor function. At each iteration, all the trajectories generated by the AMS algorithm (the reactive and non-reactive ones, with the appropriate weights\cite{Brehier2016}) are divided into non-overlapping successive sub-trajectories of $L$ timesteps, that are used in the empirical loss function \eqref{loss_ito}. Let us emphasize that this procedure actually defines empirically the measure $\mu$ (see Equation~\eqref{minimization_ito_lemma}) of the initial conditions of the trajectories of length $L$.
The minimization of \eqref{loss_ito} yields a new approximation of the forward committor, which is then used as a new importance function for the next couple of AMS runs.

Let us finally make precise the stopping criterion of the iterative procedure. After a few iterations of the proposed approach, once some reactive trajectories have been sampled, an ensemble of $n_\mathrm{convergence}$ points is chosen randomly within the configurations visited in the ensemble of sampled reactive trajectories. This ensemble of points is kept fixed. From one iteration to another, the approximate committor functions evaluated on this set of points are compared using a linear regression. This regression is actually done on the logarithm of the approximated committor values ${\chi}^\mathrm{NN}$ and on the logarithm of $1 - {\chi}^\mathrm{NN}$, to better evaluate the variation of the neural network approximation of the committor in the regions close to $R$ and $P$ respectively. See section~\ref{sec:results_ZP} for details on the architecture of neural networks. If the regression error is small between two iterations, the iterative procedure is stopped. 

\section{Numerical results}
\label{sec:num_results}

We first compare in Section~\ref{sec:results_MB} the results obtained with methodologies already existing in the literature and the new loss~\eqref{loss_ito} introduced in Section~\ref{sec:ito_loss_continuous} and ~\ref{sec:ito_loss_discretized}. We first describe the neural networks used, how the datasets are built and what kind of metrics is used to compare the performances of the various methodologies. In a second time, the methods requiring a long trajectory sampling correctly the equilibrium measure of the process, namely the loss function based on the minimization problems~\eqref{variationnal_problem_squarred_grad},~\eqref{minimization_fixed_point_canonical} and~\eqref{eq:minimization_fixed_point_roux}, are compared to the proposed method~\eqref{minimization_ito_lemma} using datasets containing the exact same number of configurations, and the initial conditions of the short trajectories are distributed according to the equilibrium measure of the dynamics~\eqref{discretized_overdamped_langevin_euler_maruyama} on the Müller--Brown potential.\cite{Muller1979} In a third time, we compare the proposed approach to the one based on minimization problem~\eqref{minimization_naive_fixed_point} with a uniform distribution of initial conditions and short trajectories. Section~\ref{sec:results_ZP} next illustrates the iterative approach described in Section~\ref{sec:ito_loss_with_ams} on a challenging 2d potential, namely the Z--potential.\cite{Frassek2021} All the results presented in the following are generated using the pytorch based code accessible at Ref.\onlinecite{TP_github}.

\subsection{Comparing the proposed method to existing ones: the Müller--Brown potential}
\label{sec:results_MB}

\subsubsection{Numerical setting}
\label{sec:MB_num_settings}

\begin{figure}[!ht]
  \centering
  \begin{subfigure}{0.53\textwidth}
         \centering
         \includegraphics[width=\textwidth]{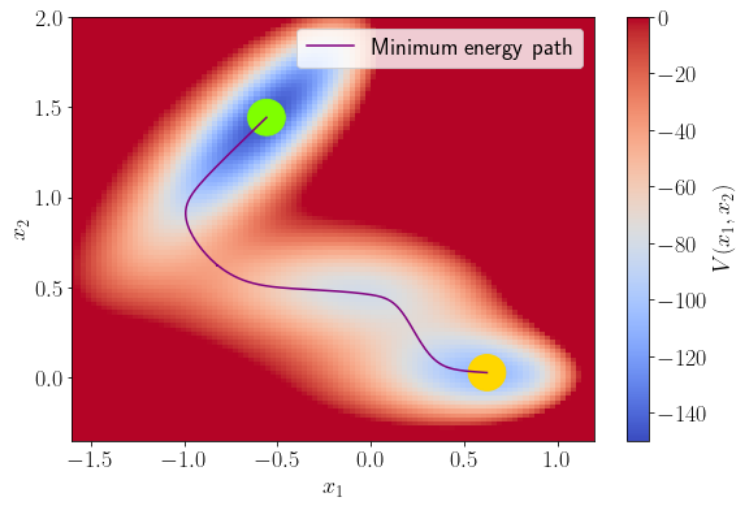}
         \caption{\label{fig:MB_pot_states} }
    \end{subfigure}
    \begin{subfigure}{0.45\textwidth}
         \centering
         \includegraphics[width=\textwidth]{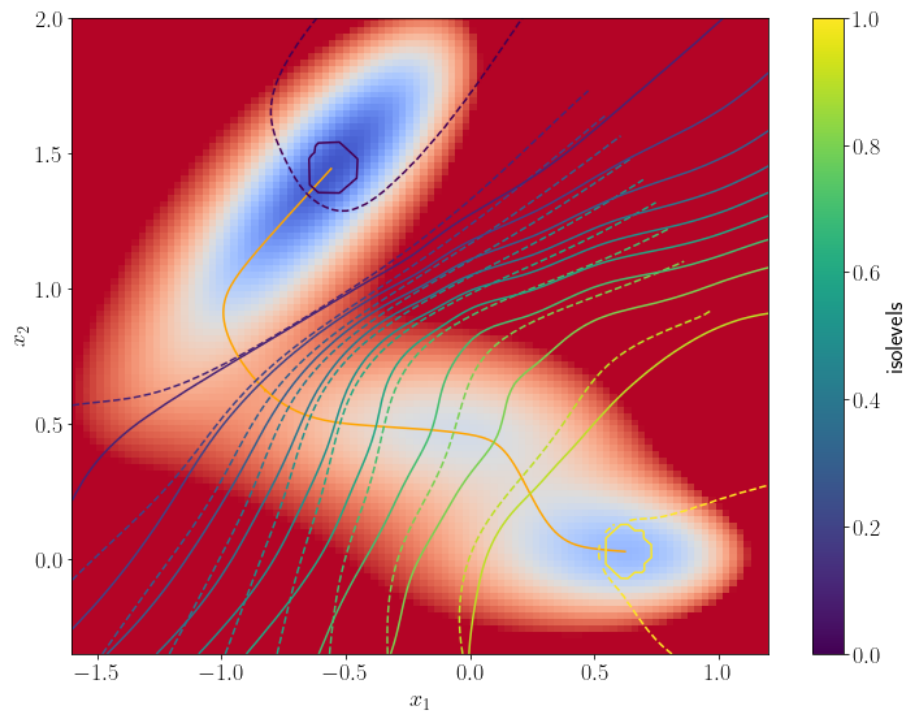}
         \caption{\label{fig:MB_committor_ex}}
     \end{subfigure}    
    \caption{\textbf{(a)} Müller--Brown potential\cite{Muller1979} on which the "reactant" $R$ and "product" $P$ states are represented as green and yellow disks, respectively. \textbf{(b)} Isolevels of a neural network committor approximation (solid line) compared with a reference finite element committor approximation (dashed line). Approximation quality (see Equations~\eqref{RMSE} and~\eqref{RMSE_log} for details): RMSE-b = $1.84 \times 10^{-2}$, RMSE-r = $4.36 \times 10^{-2}$, log-RMSE-b = $2.14$, log-RMSE-r = $0.47$.}
  \label{fig:MB_pot}
\end{figure}

We test the interest of the new loss function~\eqref{loss_ito} for various distributions of initial configurations. The test case is the 2D Müller--Brown potential:\cite{Muller1979}
\begin{equation}
\label{eq:Muller-Brown}
V(x_1,x_2) = \sum_{i=1}^{4} A_i \exp\left( a_i \left(x_1 - u_i\right)^2 + b_i \left(x_1 - u_i\right)\left(x_2 - v_i\right) + c_i \left(x_2 - v_i\right)^2 \right),
\end{equation}
with the parameters~$A = (-200, -100, -170, 15)$, $a = (-1, -1, -6.5, 0.7)$, $b = (0, 0, 11, 0.6)$, $c = (-10, -10, -6.5, 0.7)$, $u = (1, 0, -0.5, -1)$ and~$v = (0, 0.5, 1.5, 1)$.
The trajectories are obtained from the Euler--Maruyama discretization~\eqref{discretized_overdamped_langevin_euler_maruyama} of the overdamped dynamics with $\beta = 0.05$ and $\Delta t = 10^{-4}$. The "reactant" and "product" states are defined as disks of radii $0.1$ centered respectively at $(-0.558, 1.442)$ and $(0.623, 0.028)$; see Figure~\ref{fig:MB_pot_states}.

\paragraph{Neural network training parameters.} The committor function is parameterized using a fully connected feedforward neural network with two hidden layers of size 20, hyperbolic tangent as activation function and sigmoid output. The training of neural networks was performed using the Adam optimizer using an early stopping procedure: the training is stopped once the minimal validation loss does not decrease for $n_\mathrm{wait}$ epochs. The patience in the early stopping procedure $n_\mathrm{wait}$ was kept constant and equal to $50$. The learning rate for the Adam optimizer was set to $0.001$ and its other parameters to their default values. 

\paragraph{Training variability.} To account for the variations due to the initialization of the weights and biases of the neural network, 10 initial networks are generated with random weights and biases. Each of them is trained to minimize the various loss functions mentioned in Sections~\ref{sec:litt_review} and~\ref{sec:ito_loss} with 10 different seeds for the mini-batch selection. This results in 100 trainings in each case. In the various results presented and discussed below, only the result of the best network (in the sense of the validation loss) are presented. Indeed, for every training, a quarter of the dataset was  dedicated to the training, another quarter was used to compute the validation error triggering the early stopping, and the remaining half of it was used to independently evaluate the loss of the resulting neural network: the returned model is the one with the smallest latter loss.

\paragraph{Generating the training dataset.} To construct the dataset, we first run a trajectory of the discretized overdamped Langevin dynamics~\eqref{discretized_overdamped_langevin_euler_maruyama} for $2 \times 10^5$ time steps. To evaluate the loss function~\eqref{loss_ito}, $K$ configurations which are not within the state $R$ or $P$ are randomly picked within the trajectory (this implicitely defined the measure $\mu$ used in the loss, see~\eqref{loss_ito}) and short trajectories of $L$ time steps are run starting from these points. Recording the positions and Gaussians along the trajectory until $R \cup P$ is reached, we obtain a dataset $(Q^k)_{0 \le k \le K}$ with initial conditions distributed according to the Boltzmann--Gibbs measure restricted to the set $\left(\overline{R} \cup \overline{P}\right)^\mathrm{c}$. 

To compare the proposed approach to the one solving~\eqref{minimization_naive_fixed_point}, we use a dataset with uniformly distributed initial conditions. It is built by first generating $K$ random configurations from the uniform distribution on $\left([-1.6, 1.2]\times[-0.35, 2]\right) \backslash (R \cup P)$, from which a short trajectory of $L$ time steps is run for every initial condition. 


\paragraph{Metrics to measure the quality of approximate committor.} The quality of the approximate committor function is assessed by comparing the neural network approximation to a reference committor computed with finite elements.\cite{sule2023sharperrorestimatestarget, Mar1akc_githum_fem} To do so, two types of errors are computed. These two errors made precise below are computed for two sets of points $\left\{ \textbf{q}_i^\mathrm{boltz} \right\}_{1 \leqslant i \leqslant N}$ and $\left\{ \textbf{q}_i^\mathrm{react} \right\}_{1 \leqslant i \leqslant N}$, with $N=10^5$, sampled respectively from the Boltzmann--Gibbs distribution sampled using a long trajectory of $10^7$ time steps of the process~\eqref{discretized_overdamped_langevin_euler_maruyama} and the reactive trajectory density sampled by 10 repetitions of the AMS algorithm (using the finite element approximation of the committor as a reaction coordinate and $N_\mathrm{rep}=200$).

The first error is the Root Mean Square Error (RMSE): 
\begin{equation}
    \label{RMSE}
    \mathrm{RMSE} = \sqrt{\frac{1}{N}\sum_{i=1}^{N}\bigg(\chi^\mathrm{FE}(\textbf{q}_i) - \chi^\mathrm{NN}(\textbf{q}_i)\bigg)^2},
\end{equation}
where $\chi^\mathrm{NN}$ and $\chi^\mathrm{FE}$ are the committor functions approximated using respectively the neural network and finite elements. To put a stronger emphasis on the error near the states~$R$ and~$P$, we also monitor the RMSE of the logarithm of the committor:
\begin{equation}
    \label{RMSE_log}
    \begin{aligned}
        \mathrm{RMSE-log} = \Bigg[ &\frac{1}{N}\sum_{i=1}^{N}\bigg(\ln{\chi^\mathrm{FE}(\textbf{q}_i)} - \ln{\chi^\mathrm{NN}(\textbf{q}_i)}\bigg)^2 \\ 
        & + \frac{1}{N}\sum_{i=1}^{N} \bigg(\ln{\left[1 - \chi^\mathrm{FE}(\textbf{q}_i)\right]} - \ln{\left[1 - \chi^\mathrm{NN}(\textbf{q}_i)\right]}\bigg)^2\Bigg]^\frac{1}{2}.
    \end{aligned}
\end{equation}

Considering the two errors~\eqref{RMSE} and~\eqref{RMSE_log} computed on each set of points $\left\{ \textbf{q}_i^\mathrm{boltz} \right\}_{1 \leqslant i \leqslant N}$ and $\left\{ \textbf{q}_i^\mathrm{react} \right\}_{1 \leqslant i \leqslant N}$, four errors are presented to compare the various methods tested to approximate the committor.  They are labeled as RMSE-b, RMSE-r, RMSE-log-b, RMSE-log-r. An illustration of isolevels for a committor approximation with given metrics is presented in Figure~\ref{fig:MB_committor_ex}.

\subsubsection{Results}
\label{sec:MB_results}

\begin{figure}[!ht]
  \centering
  \begin{subfigure}{0.44\textwidth}
         \centering
         \includegraphics[width=\textwidth]{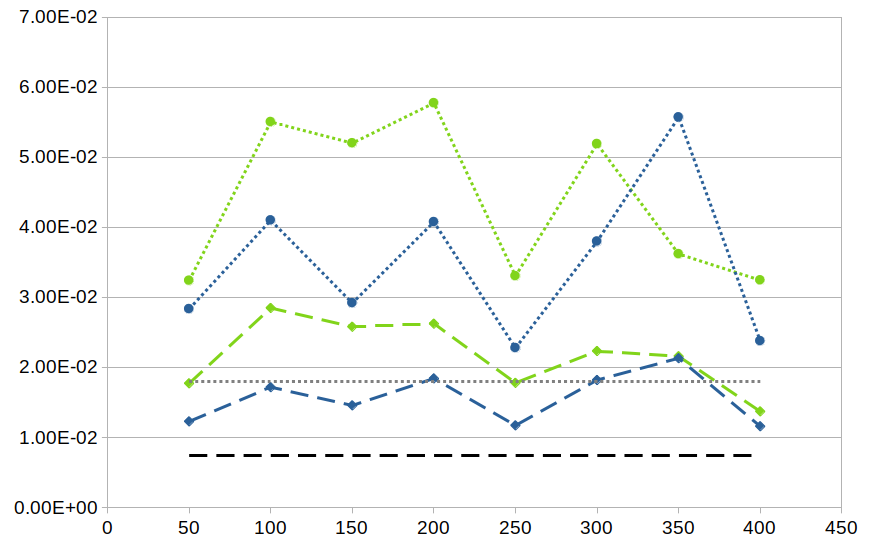}
         \caption{\label{fig:err_boltz_litt} }
    \end{subfigure}
    \begin{subfigure}{0.44\textwidth}
         \centering
         \includegraphics[width=\textwidth]{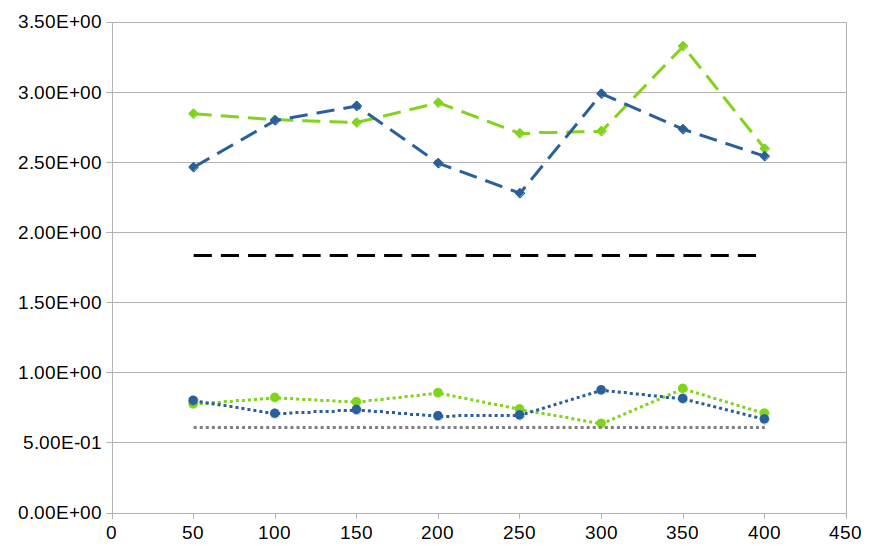}
         \caption{\label{fig:err_boltz_litt_log}}
     \end{subfigure}    
    \caption{Errors on the approximate committor function
    as a function of the time-lag $\vartheta$ expressed in number of time steps  (when relevant), trained for various loss functions:  \eqref{variationnal_problem_squarred_grad} with $\alpha=100$ (black, no dependency on $\vartheta$), \eqref{minimization_fixed_point_canonical} (blue) and \eqref{eq:minimization_fixed_point_roux} (green). Fig. \textbf{(a)} RMSE-b: diamonds and dashed lines; RMSE-r: circles and dotted lines; Fig. \textbf{(b)} log-RMSE-b: diamonds and dashed lines; log-RMSE-r: circles and dotted lines.}
  \label{fig:erreur_boltz_litt}
\end{figure}

\paragraph{Comparison of methods using long trajectory sampling the equilibrium measure.} First, we compare the methods based on the minimization problems~\eqref{variationnal_problem_squarred_grad},~\eqref{minimization_fixed_point_canonical} and~\eqref{eq:minimization_fixed_point_roux} to the proposed method~\eqref{minimization_ito_lemma}. The train and validation dataset was composed of $1.6 \times 10^4$ points and the batch size was fixed to $100$. The first loss tested in this setting is~\eqref{variationnal_problem_squarred_grad} (PDE variational loss, with penalized boundary conditions), with $\alpha=100$. Other losses based on the transition operator (see~\eqref{minimization_fixed_point_canonical} or~\eqref{eq:minimization_fixed_point_roux}) are tested with various choices of time-lag. From Figure~\ref{fig:erreur_boltz_litt}, we observe that the variational approach based on~\eqref{variationnal_problem_squarred_grad} is slightly better than the approaches based on the transition operator.

\begin{figure}[!ht]
  \centering
  \begin{subfigure}{0.44\textwidth}
         \centering
         \includegraphics[width=\textwidth]{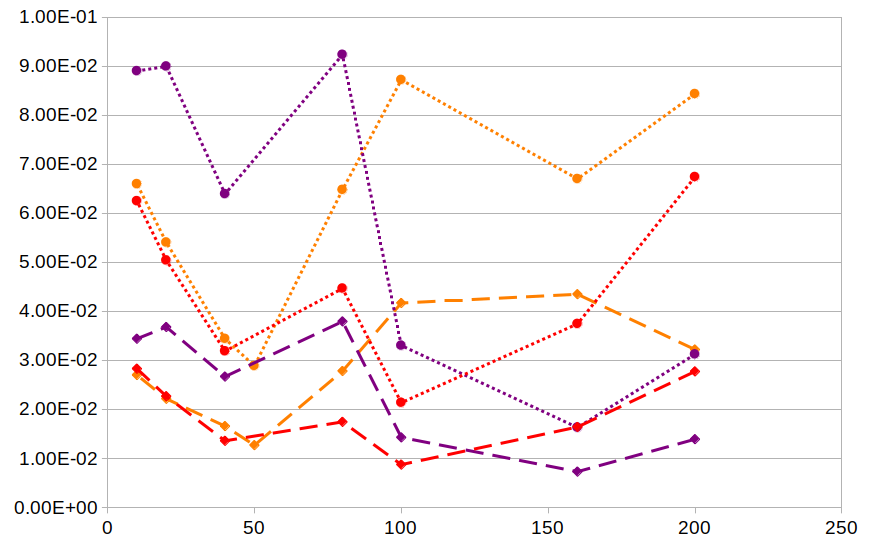}
         \caption{\label{fig:err_boltz_ito} }
    \end{subfigure}
    \begin{subfigure}{0.44\textwidth}
         \centering
         \includegraphics[width=\textwidth]{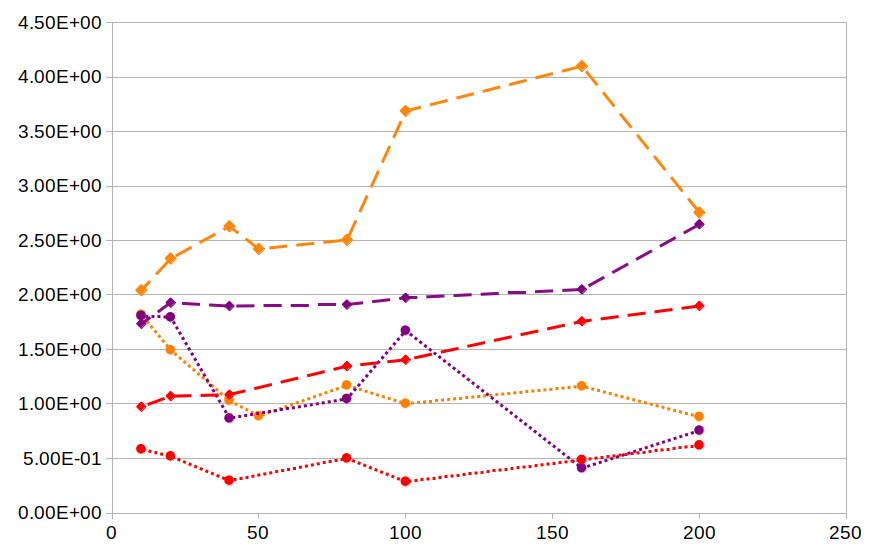}
         \caption{\label{fig:err_boltz_ito_log}}
     \end{subfigure}    
    \caption{Errors on the approximate committor function
    as a function of the time-lag $\vartheta$ expressed in number of time steps, trained for various loss functions:~\eqref{loss_ito} with $h=\mathrm{Id}$ (orange), \eqref{loss_ito} with $h=\ln$ and $\varepsilon=0.1$ (purple), and \eqref{loss_ito} with $h=\ln$ and $\varepsilon=0.01$ (red). Fig. \textbf{(a)} RMSE-b: diamonds and dashed line RMSE-r: circles and dotted lines; Fig. \textbf{(b)} log-RMSE-b: diamonds and dashed line log-RMSE-r: circles and dotted lines.}
  \label{fig:erreur_boltz_ito}
\end{figure}

To compare these reference results with the method we propose in this work, we test various losses based on Equation~\eqref{loss_ito}. The total dataset size is equal to $KL$ and the batch size is fixed to $10$. To compare the methods on an equal footing, we make sure that we use exactly the same number of training configurations to obtain these reference results as in the training of the previous methods: the total dataset size $K \times L = 1.6 \times 10^4$ is kept fixed for the various tests. The results are presented on Figure~\ref{fig:erreur_boltz_ito}.

Comparing the results in Figures~\ref{fig:erreur_boltz_litt} and~\ref{fig:erreur_boltz_ito}, we observe that  the results obtained with the loss function~\eqref{loss_ito} are slightly better than those obtained with the loss functions~\eqref{variationnal_problem_squarred_grad},~\eqref{minimization_fixed_point_canonical} and~\eqref{eq:minimization_fixed_point_roux} using the same number of configurations in the training dataset.

Considering the test cases with  $h = \ln$ and various values of $\varepsilon$, it seems that putting more emphasis on configurations close to the $R$ and $P$ sets has a positive impact. Indeed, $h = \ln$ leads to smaller values of all the monitored errors thant $h=\mathrm{Id}$. 

\begin{figure}[!ht]
  \centering
  \begin{subfigure}{0.44\textwidth}
         \centering
         \includegraphics[width=\textwidth]{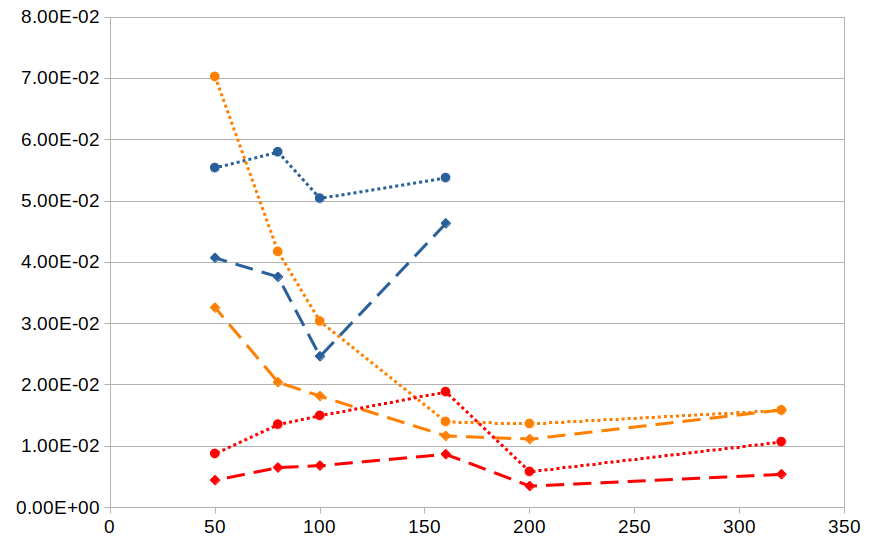}
         \caption{\label{fig:err_unif_litt_et_ito} }
    \end{subfigure}
    \begin{subfigure}{0.44\textwidth}
         \centering
         \includegraphics[width=\textwidth]{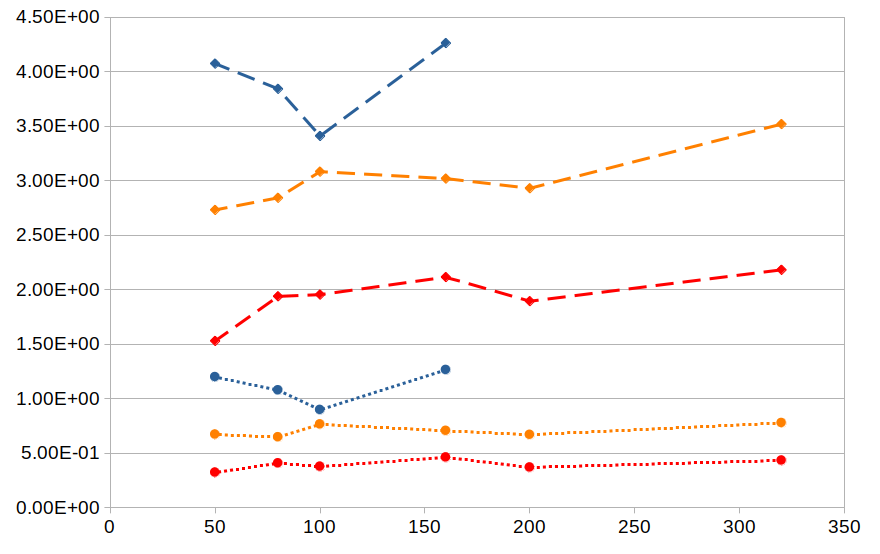}
         \caption{\label{fig:err_unif_litt_et_ito_log}}
     \end{subfigure}    
    \caption{Errors on the approximate committor function
    as a function of the time-lag $\vartheta$ expressed in number of time steps, trained for various loss functions: \eqref{minimization_naive_fixed_point} (blue); ~\eqref{loss_ito} with $h=\mathrm{Id}$ (orange) and \eqref{loss_ito} $h=\ln$ with $\varepsilon=0.01$ (red). Fig. \textbf{(a)} RMSE-b: diamonds and dashed lines RMSE-r: circles and dotted lines; Fig. \textbf{(b)} log-RMSE-b: diamonds and dashed lines  log-RMSE-r: circles and dotted lines.}
  \label{fig:erreur_unif_litt_et_ito}
\end{figure}

\paragraph{Comparison of methods using uniformly distributed samples.} We next compare the results obtained with the newly introduced loss~\eqref{loss_ito} with the results obtained with the loss~\eqref{minimization_naive_fixed_point} (see Ref.~\onlinecite{STRAHAN2023} and Section~\ref{sec:residuals_propagator}). In both case we set $\mu$ to be the uniform distribution on $\left(\overline{R} \cup \overline{P} \right)^c$. The total dataset size is equal to $KL = 1.6 \times 10^5$ and the batch size was fixed to $100$, K being the number of initial conditions and~\eqref{loss_ito}. For~\eqref{minimization_naive_fixed_point}, one needs to run two trajectories starting from each initial condition to estimate the expectation, while in~\eqref{loss_ito}, only one trajectory is needed. In Figure~\ref{fig:erreur_unif_litt_et_ito}, we compare the results of models trained using dataset requiring exactly the same number of force calls. In this setting, it is clear that the approach based on~\eqref{minimization_naive_fixed_point} leads to less accurate estimations. This is likely due to the fact that only the first and last configurations of a trajectory are used to approximate the committor function, while the approach using the residual of It{\=o}'s formula uses all the configurations to train the approximate committor. 

\subsection{Iterative learning of the committor function with AMS}
\label{sec:results_ZP}
The objective of this section is to test numerically the iterative procedure described in Section~\ref{sec:ito_loss_with_ams}.

\subsubsection{Numerical settings}
\label{sec:ZP_num_settings}

\begin{figure}[!ht]
  \centering
    \includegraphics[width=\textwidth]{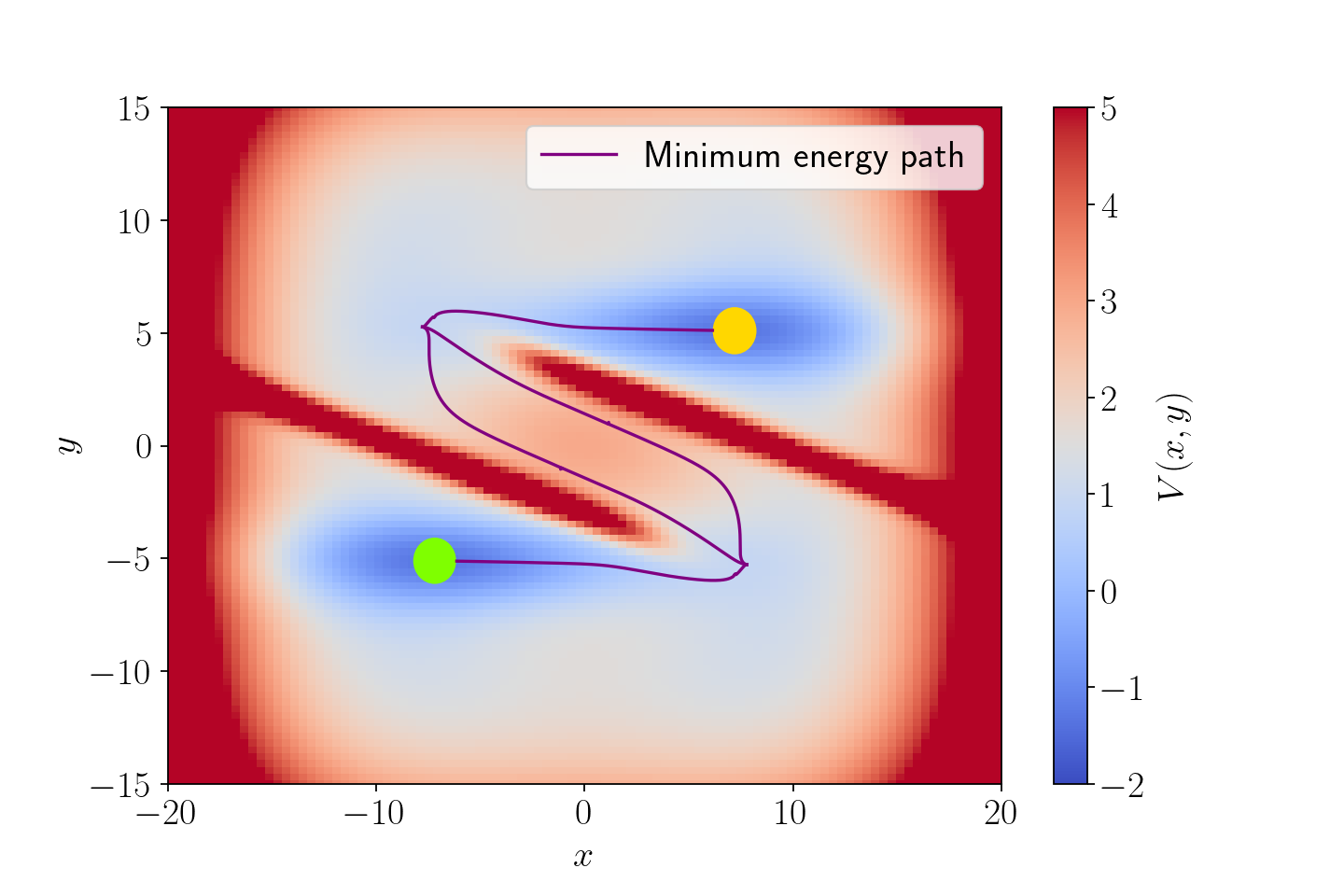}
    \caption{Z--potential.\cite{Frassek2021} The reactant $R$ and product $P$ states are the green and yellow discs, respectively. The purple curve correspond to the minimum energy paths (MEPs) linking the states $R$ and $P$. The first MEP goes from the Reactant to the second local minimum (around (7, -5)), then two parallel MEPs link this local minimum to its symmetrically (with respect to the origin (0, 0)) equivalent, and finally the last MEP links this local minimum to the product basin $P$.}
  \label{fig:ZP_pot}
\end{figure}

\paragraph{Z--Potential.} We consider the Z--potential,\cite{Frassek2021} represented on Figure~\ref{fig:ZP_pot}:
\begin{equation}
\label{eq:Z_potential}
\begin{aligned}
	V(x,y) = &- 3 \mathrm{e}^{-0.01\left(x + 5\right)^2 - 0.2\left(y + 5\right)^2} - 3 \mathrm{e}^{-0.01\left(x - 5\right)^2 - 0.2\left(y - 5\right)^2}  + 3\mathrm{e}^{-0.01\left(x^4 + y^4\right)} \\ & + \frac{5\mathrm{e}^{-0.2\left(x + 3\left(y+3\right)\right)^2}}{1 + \mathrm{e}^{x -3}} + \frac{5\mathrm{e}^{-0.2\left(x + 3\left(y-3\right)\right)^2}}{1 + \mathrm{e}^{-x -3}}  + \frac{x^2 + y^2}{20480} .
\end{aligned}
\end{equation}
It is not easy for an automatic procedure to approximate the committor for such a system since the transition path between the two global minima has a Z shape, which cannot be obtained by a simple linear interpolation between the minima.

\paragraph{Dynamics and state definitions.} The system evolves according to the overdamped Langevin dynamics with $\beta = 3$, discretized with the Euler--Maruyama scheme~\eqref{discretized_overdamped_langevin_euler_maruyama} with $\Delta t = 0.05$. The "reactant" and "product" states $R$ and $P$ are defined as discs of radii $0.98$ around the points $(-7.20, -5.10)$ and $(7.20, 5.10)$. We define circles with the same centers and a radius of $1.0$  as $\Sigma_R$ and $\Sigma_P$. The initial conditions for AMS are the first configuration visited by the dynamics outside these circles after leaving the respective states, see~SI.6 for details. 

\paragraph{Neural network training parameters. } All the approximate committor functions are parameterized by a fully connected feed-forward neural network with architecture (2,~20,~20,~1), hyperbolic tangent activation functions and a sigmoid output. The model is trained using the Adam optimizer with a learning rate of $0.001$ and its other parameters set to their default values that can be found in PyTorch documentation. The early stopping criterion is fixed to $n_\mathrm{wait} = 50$ and the batch size to $100$ unless otherwise stated. The length of trajectories to build the training dataset is $L=400$. This choice of length $L$ is motivated by the fact that reactive trajectories have an average length around $1000\Delta t$, thus the length $L$ corresponds roughly to the time to commit to one of the two basins of the potential when starting from the transition region. Finally, the loss function used was the one with $h=\ln$ and $\varepsilon=10^{-5}$.

\subsubsection{Results}
\label{sec:ZP_results}

Let us first present a short overview of the obtained results during the iterative procedure applied to the Z--potential; a more detailed discussion follows.

\begin{figure}[!ht]
  \centering
    \begin{subfigure}[b]{0.44\textwidth}
         \centering
         \includegraphics[width=\textwidth]{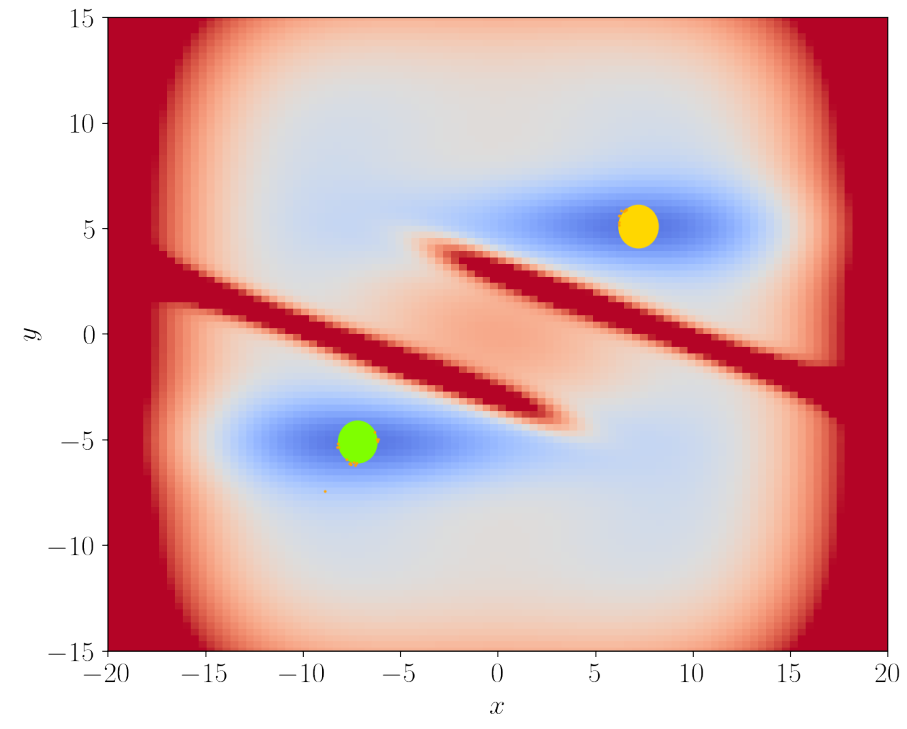}
         \caption{\label{fig:first_ZP_training_b}}
    \end{subfigure}
    \begin{subfigure}[b]{0.54\textwidth}
         \centering
         \includegraphics[width=\textwidth]{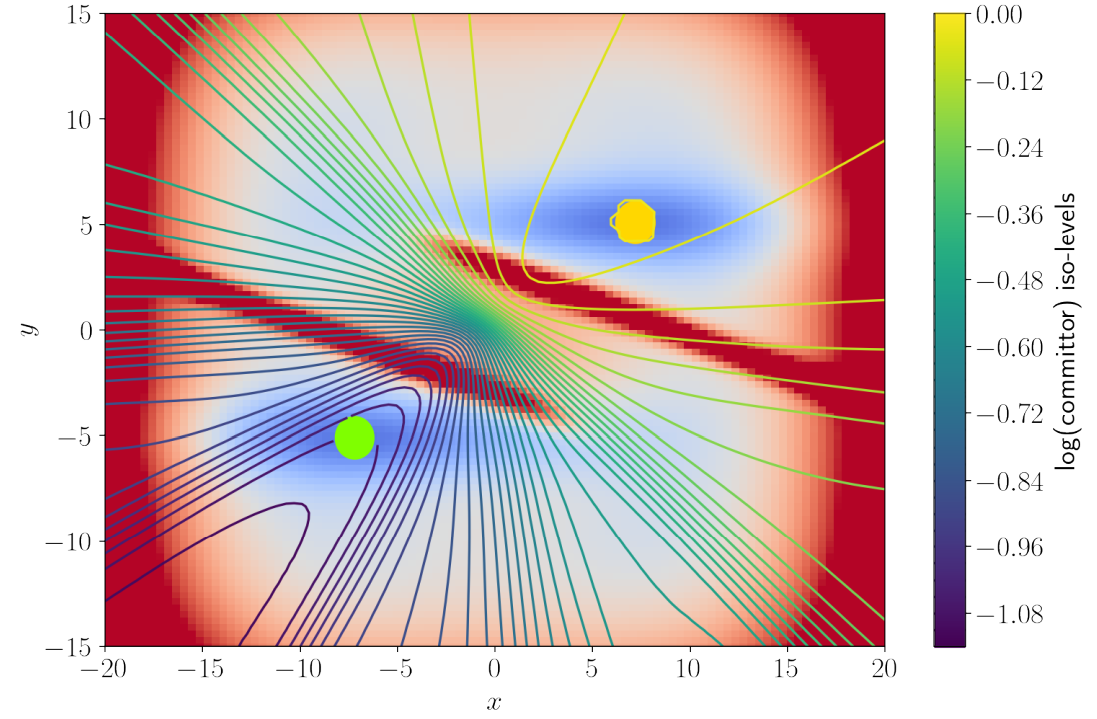}
         \caption{\label{fig:first_ZP_training_a}}
     \end{subfigure}
  \caption{Result at iteration 0. (a) Initial conditions of the trajectories in the training dataset (b) Isolevels of the logarithm of the approximate committor function trained using~\eqref{loss_ito}.}
  \label{fig:first_ZP_training}
\end{figure}

\begin{figure}[!ht]
  \centering
     \begin{subfigure}[b]{0.44\textwidth}
         \centering
         \includegraphics[width=\textwidth]{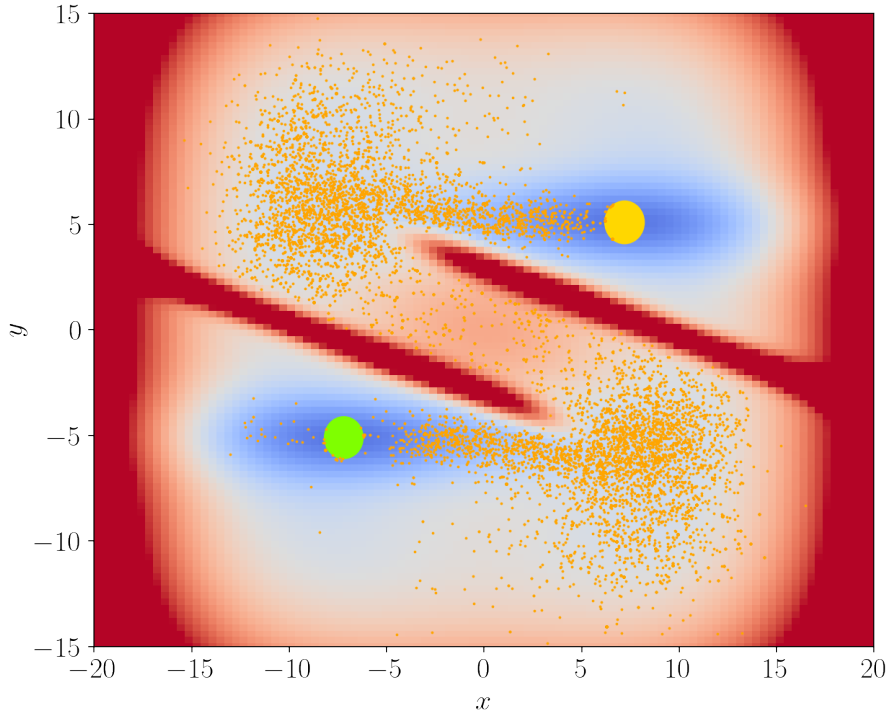}
         \caption{\label{fig:second_ZP_training_b}}
     \end{subfigure}
     \begin{subfigure}[b]{0.54\textwidth}
         \centering
         \includegraphics[width=\textwidth]{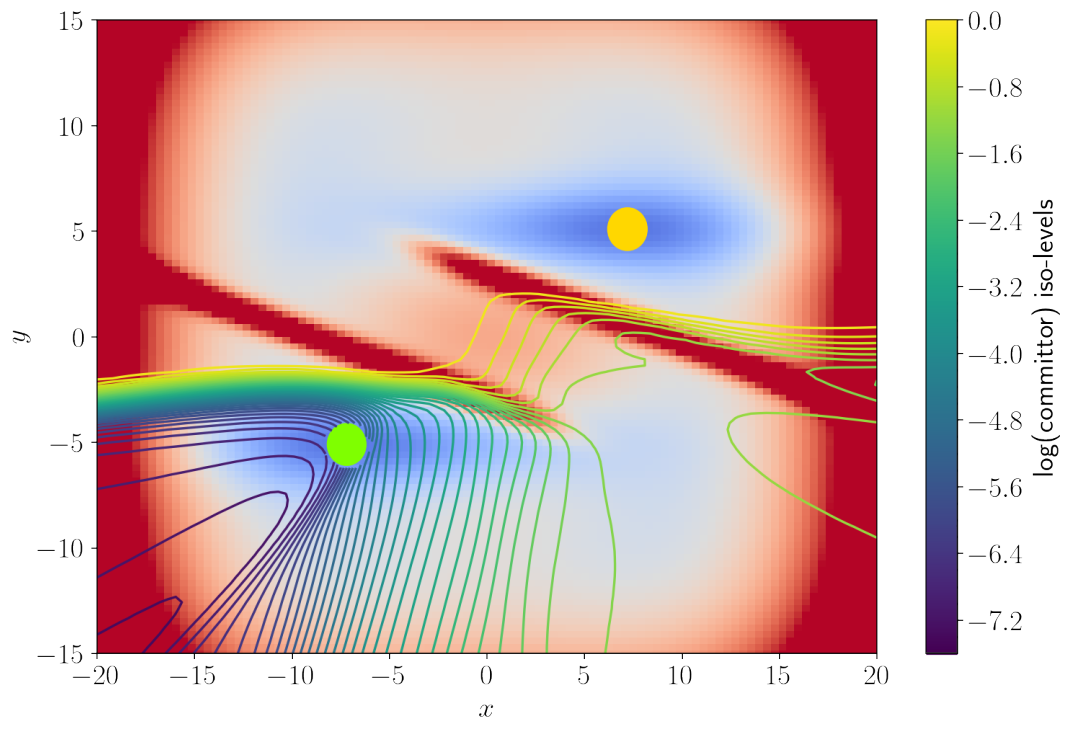}
         \caption{\label{fig:second_ZP_training_a}}
     \end{subfigure}
  \caption{Results at iteration 1. (a) Initial conditions of the trajectories in the training dataset. (b) Isolevels of the logarithm of the approximate committor function.}
  \label{fig:second_ZP_training}
\end{figure}

\begin{figure}[!ht]
  \centering
     \begin{subfigure}[b]{0.44\textwidth}
         \centering
         \includegraphics[width=\textwidth]{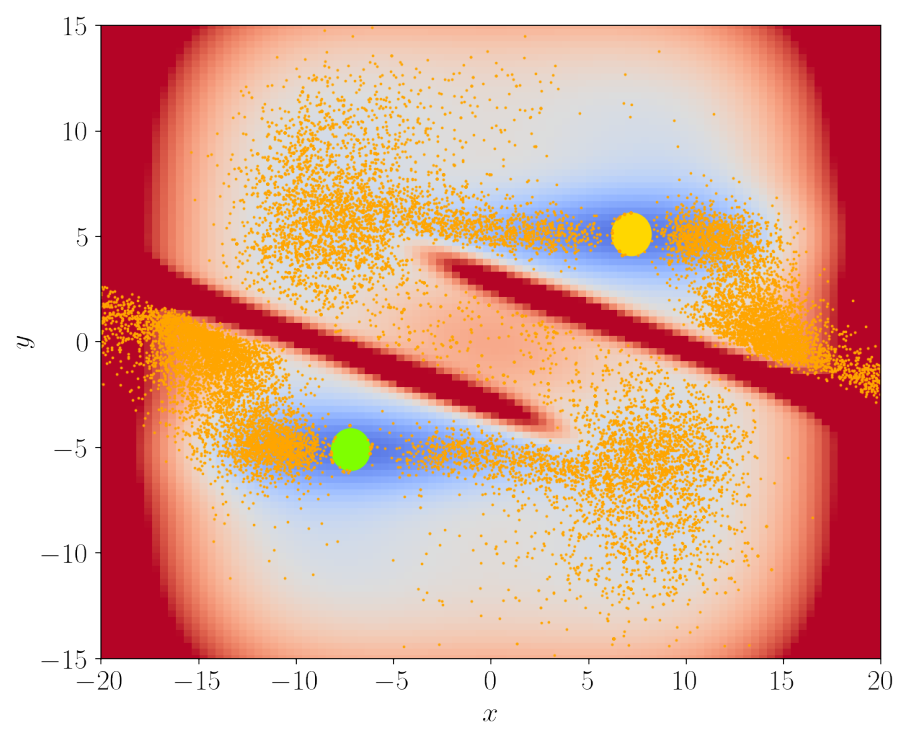}
         \caption{\label{fig:third_ZP_training_b}}
     \end{subfigure}
     \begin{subfigure}[b]{0.54\textwidth}
         \centering
         \includegraphics[width=\textwidth]{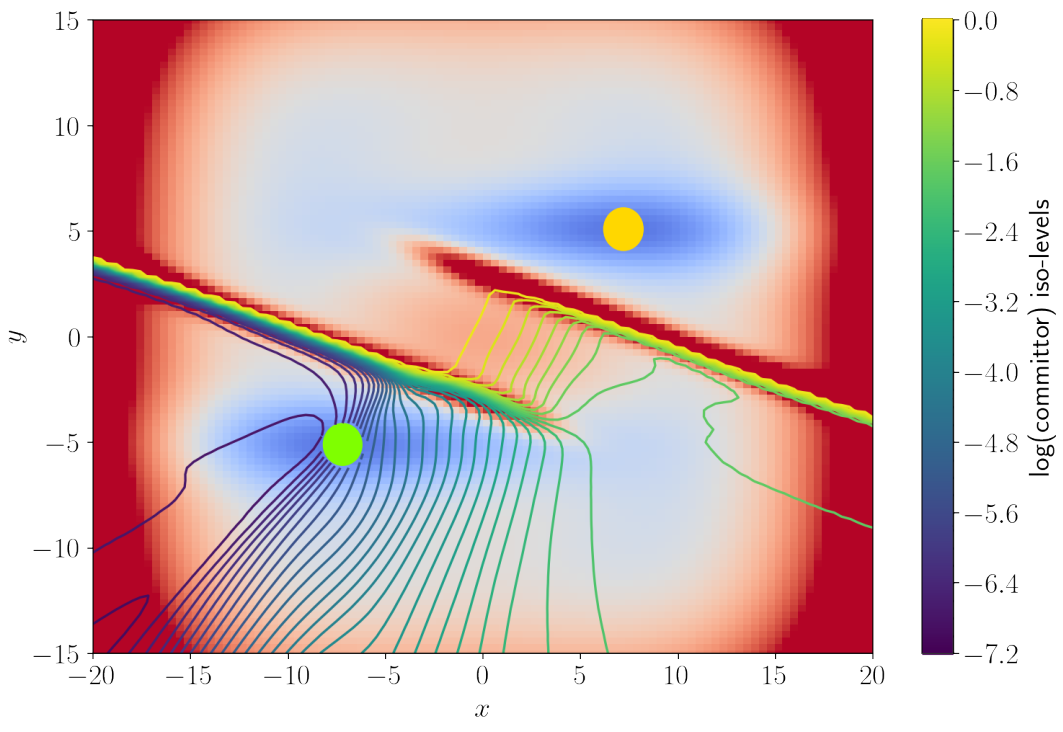}
         \caption{\label{fig:third_ZP_training_a}}
     \end{subfigure}
  \caption{Results at iteration 2. (a) Initial conditions of the trajectories in the training dataset. (b) Isolevels of the logarithm of the approximate committor function.}
  \label{fig:third_ZP_training}
\end{figure}

\paragraph{Overview.} The Figures~\ref{fig:first_ZP_training} to~\ref{fig:third_ZP_training}  present the evolution of the approximation of the committor function. In all the plots, the initial points of the subtrajectories of length $L$ used in the loss~\eqref{loss_ito} are represented on the subfigures (a) while the subfigures (b) represent the logarithm of the (approximated) committor. The points represented on the subfigures (a) progressively cover the reactive path as well as the vicinities of the states $R$ and $P$ which translates in to better and better quality of approximated committor as represented in the subfigures (b). The iterative procedure converges after 4 iterations.

\paragraph{Detailed description.} The training procedure starts with the sampling of 20 initial conditions for a forward (on $\Sigma_R$) and a backward (on $\Sigma_P$) run of AMS. With these sampled trajectories, a dataset of $K=20$ trajectories of length $L=400$ is used to minimize the loss~\eqref{loss_ito} with a batch size of 10, which leads to a first approximation of the committor. In that particular case, as the dataset is small, the training is not carried out until early stopping but stopped after 100 epochs. The initial conditions of the trajectories and the isolevels of the log of this first approximate committor function are represented on Figure~\ref{fig:first_ZP_training}.
Using this approximate committor function as reaction coordinate in AMS, a backward and forward AMS are run with 20 replicas each. This concludes iteration 0. 

The procedure then proceeds by successively computing an approximation of the committor function, and running new AMS algorithms (20 runs forward and backward). At each iteration, the dataset used to approximate the committor function is built as follows. Using all the sampled reactive and non-reactive trajectories, the training dataset of the approximate committor function in increased. More precisely, by cutting each AMS replica's trajectory (reactive or not) into sub trajectories of length $400 \Delta t$, we obtain a dataset made of trajectories with initial conditions covering progressively the configurations along the reactive paths and in the vicinity of $R$ and $P$ (Figures~\ref{fig:first_ZP_training_b},~\ref{fig:second_ZP_training_b} and~\ref{fig:third_ZP_training_b}). The weight of each sub-trajectory built by cutting a replica trajectory is the weight of the replica computed during the AMS simulation.\cite{Brehier2016}

\begin{figure}[!ht]
  \centering
     \begin{subfigure}[b]{0.45\textwidth}
         \centering
         \includegraphics[width=\textwidth]{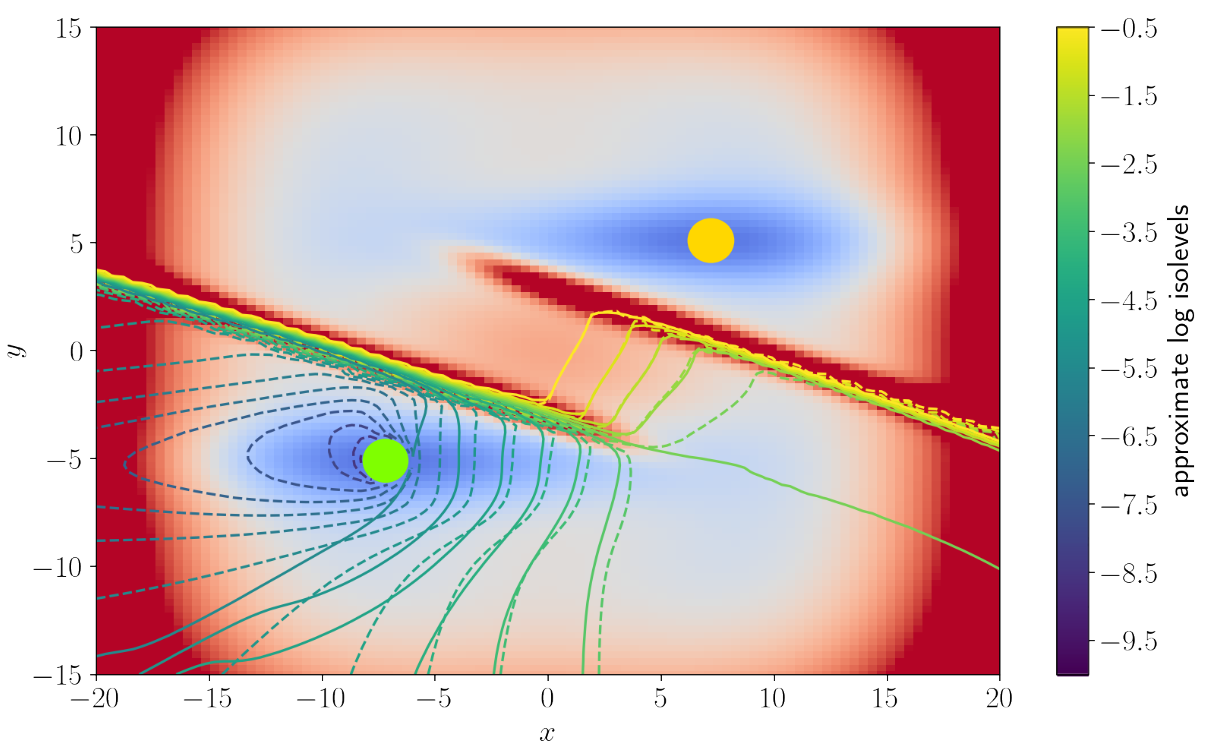}
         \caption{\label{fig:fifth_ZP_training_a}}
     \end{subfigure}
     \begin{subfigure}[b]{0.45\textwidth}
         \centering
         \includegraphics[width=\textwidth]{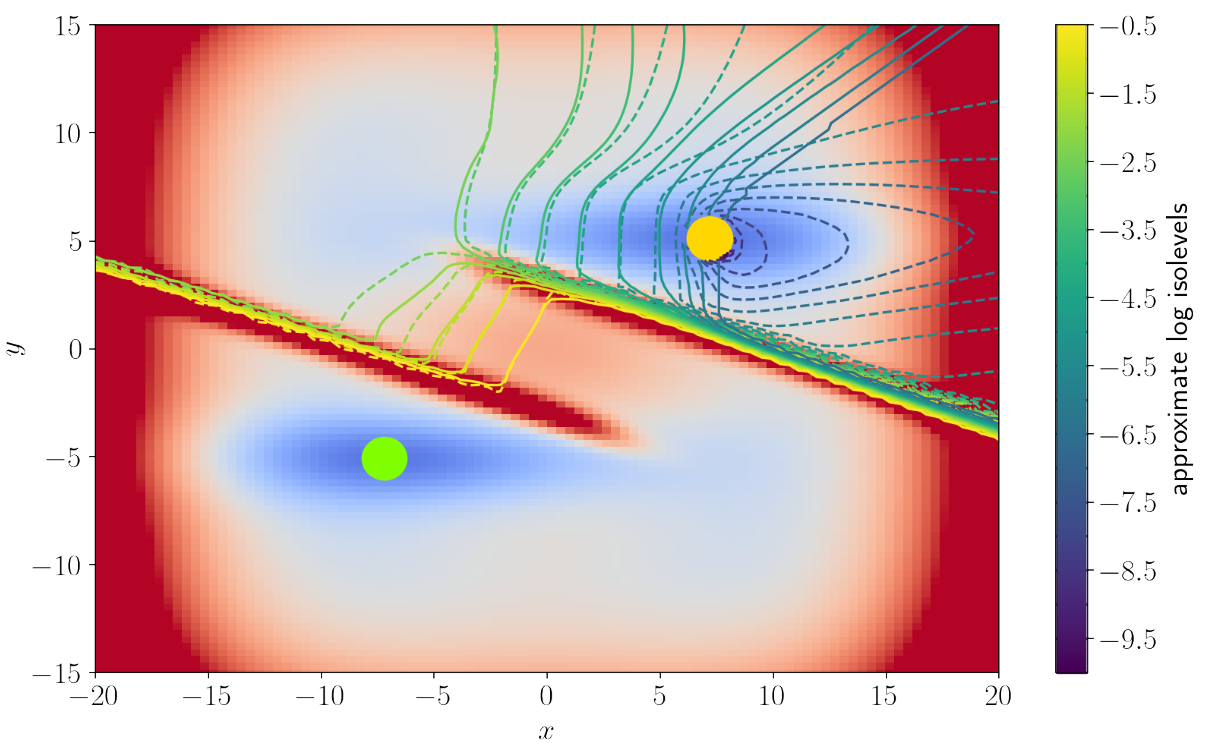}
         \caption{\label{fig:fifth_ZP_training_b}}
     \end{subfigure}
  \caption{Best model obtained after the iterative training procedure using AMS. (a) (and (b)) Isolevels of the logarithm of the approximate committor (or the reverse $P \to R$ committor) (full lines) and finite elements approximation (dotted lines)}
  \label{fig:fifth_ZP_training}
\end{figure}

The converged approximation of the committor function is presented on Figure~\ref{fig:fifth_ZP_training}; the final approximate committor function is obtained after 4 iterations including AMS simulations. The final training dataset is composed of  $60 \times 10^3$ trajectories. The configurations to stop the procedure were taken within the ensemble of reactive trajectories configurations sampled during the $3^\mathrm{rd}$ iteration. The linear regression between the log of the approximate committor (and the log of the reverse $P \to R$ committor) between iterations 3 and 4 is 0.998. 

Finally, we next evaluate whether this approximate committor function allows to obtain better results with AMS than a naive reaction coordinate such as   $$\xi_\mathrm{interp}(x,y) = \frac{\left(x_\mathrm{P} - x_\mathrm{R}\right)\left(x - x_\mathrm{R}\right) + \left(y_\mathrm{P} - y_\mathrm{R}\right)\left(y - y_\mathrm{R}\right)}{\sqrt{\left(x_\mathrm{P} - x_\mathrm{R}\right)^2 + \left(y_\mathrm{P} - y_\mathrm{R}\right)^2}}$$ which linearly interpolates between $R$ and $P$. 
We run to this end 10 forward and backward AMS with 100 replicas to estimate the averaged committor probability on the exit set of $R$ and the empirical standard deviation. This procedure is repeated using the $\xi_\mathrm{interp}$ coordinate mentioned above, the approximate committor function obtained with the procedure detailed above and the committor obtained with finite elements method. We observe that the $\xi_\mathrm{interp}$ reaction coordinate leads to confidence intervals that contains 0, which for probability is nonphysical. Moreover, the target probability is not in the empirical confidence interval: this is an example of the apparent bias phenomenon which demonstrates that $\xi_\mathrm{interp}$ is a really bad score function.\cite{Glasserman1998, Brehier2015}

\begin{table}[!ht]
     \centering
     \caption{95\% confidence interval of the transition probability estimated with 100 forward and backward AMS runs using various reaction coordinates. "neural network committor" stands for the approximation of the committor obtained at the end of the iterative procedure and "FE committor" is the finite element approximation of the committor function. }
	\label{confidence_interval_proba_various_rcs}
     \begin{tabular}{c|c|c|c}
      RC &  $\xi_\mathrm{interp}$ & neural network committor & FE committor\\
      \hline
      \hline
      Forward & & & \\
      $p \pm \frac{1.96}{\sqrt{10}} \, \sigma_p$ & $(3.39 \pm 6.58) \times 10^{-8}$ & $(4.41 \pm 1.07) \times 10^{-7}$ & $(5.58 \pm 1.02) \times 10^{-7}$ \\ 
      \hline
      \hline
      Backward & & & \\
      $p \pm \frac{1.96}{\sqrt{10}} \, \sigma_p$ & $(0.89 \pm 1.52) \times 10^{-8}$ & $(6.57 \pm 1.48) \times 10^{-7}$ & $(5.09 \pm 1.) \times 10^{-7}$ \\
     \end{tabular}
 \end{table}

\section{Conclusion}

The committor function is a high dimensional function quantifying the probability that a configuration reacts. It is thus an optimal choice of reaction coordinate to describe reactive processes in chemistry and biology.\cite{VandenEijnden2005, Brehier2015} We introduced a novel methodology based on the minimization of the residuals of the It\={o}'s formula applied to the committor function to approximate it by a neural network. We compared this approach to established ones on the particular case of the Müller--Brown potential. Even though we did not fully explore the hyper parameter space of established methodologies (but used reasonable values for these hyperparameters), on this test case, with similar amount of data, the proposed methodology provides approximations of the same quality. Extending our approach to the logarithm of the scaled and shifted committor, we observed that this formulation allows to put some emphasis on regions of the configuration space closer to reactant and product states, leading to higher quality approximations. We quickly checked that this approach can also be extended to biased dynamics which allows faster exploration of the relevant parts of the potential energy surface. Finally, we included this approach in an active learning scheme in which the AMS algorithm is used to enhance the sampling of reactive trajectories, thus allowing to quickly construct a dataset of trajectories from which an accurate committor function approximation can be built. In this procedure, the reaction coordinate used for AMS is the approximate committor function which in principle allows to address reactivity problems given only the definitions of the reactant and product states. 

A future perspective of this work is to extend the newly proposed approach to underdamped Langevin dynamics, which would allow to define reaction coordinates that also depends on momenta. Another perspective would be to implement such committor approximation as a secondary output of a neural network interatomic potential. This would allow to take advantage of the invariances of the existing network while minimizing the computational cost in inference mode. 

\begin{acknowledgments}
The authors thank Hadrien Vroyland, Christophe Chipot and Benoit Roux for fruitful discussions. Part of this research was performed while TP and GS were visiting the Institute for Mathematical and Statistical Innovation (IMSI), which is supported by the National Science Foundation (Grant No. DMS-1929348). The Agence Nationale de la Recherche under France 2030 is acknowledged for support in the context of the MAMABIO project (contract ANR-22-PEBB-0009, B-BEST PEPR).
\end{acknowledgments}

\bibliography{committor}

\end{document}



\title{Approximating committor functions: \\ Objective functions and enhanced sampling \\ Supplementary Information} 



\author{Thomas Pigeon}
    \email{thomas.pigeon@ifpen.fr}
    \affiliation{IFP Energies Nouvelles, Rond-Point de l’Echangeur de Solaize, BP 3, 69360 Solaize, France}

\author{Gabriel Stoltz}
    \email{gabriel.stoltz@enpc.fr}
    \affiliation{CERMICS, École nationale des ponts et chaussées, Institut Polytechnique de Paris, 6-8 Avenue Blaise Pascal, 77455, Marne-la-Vallée, France}
    \altaffiliation[Also at ]{MATHERIALS team-project, Inria Paris, 48 Rue Barrault, 75013 Paris, France}

\author{Tony Lelièvre}
    \email{tony.lelievre@enpc.fr}
    \affiliation{CERMICS, École nationale des ponts et chaussées, Institut Polytechnique de Paris, 6-8 Avenue Blaise Pascal, 77455, Marne-la-Vallée, France}
    \altaffiliation[Also at ]{MATHERIALS team-project, Inria Paris, 48 Rue Barrault, 75013 Paris, France}

\date{\today}

\pacs{}

\maketitle 

\section{Proof that the committor solves the minimization problem~(14}
\label{sec:proof_variationnal_PDE}

We consider the minimization problem~(14) of the main text, that we rewrite for convenience: 
\begin{equation}
\label{variationnal_problem_squarred_grad}
    \underset{f}{\mathrm{inf}} \left\{ \int_{\left(\overline{R} \cup \overline{P}\right)^\mathrm{c}} \left|\nabla f(\textbf{q}) \right|^2 \mathrm{e}^{-\beta V(\textbf{q})} \, \mathrm{d}\textbf{q} \, \middle| \, f(\textbf{q)} = \mathbb{1}_P(\textbf{q}), \, \forall \, \textbf{q} \in \left(\overline{R} \cup \overline{P}\right) \right\}.
\end{equation} 
Notice that the problem~\eqref{variationnal_problem_squarred_grad} is for example well posed for functions in the Sobolev space $H^1 (\exp(-\beta V)/Z)$. Let us show, following  Ref. \onlinecite{Khoo2018}, that solutions to the minimization problem~\eqref{variationnal_problem_squarred_grad} satisfy the equation:
\begin{equation}
\label{back_kolmo_committor}
\left\{
\begin{aligned}
    &\mathcal{L}_{\mathrm{ovd}} \chi(\textbf{q}) = 0, \qquad \forall \, \textbf{q} \in \left(\overline{R} \cup \overline{P}\right)^\mathrm{c} ,\\
    & \chi (\textbf{q}) = 0, \quad \textbf{q} \in \overline{R} \; ; \qquad \chi (\textbf{q}) = 1, \quad \textbf{q} \in \overline{P}.
\end{aligned}
\right.
\end{equation}
To this end, we write $f$ as $f_\lambda(\textbf{q}) = f^*(\textbf{q}) + \lambda \delta f(\textbf{q})$ where $f^*$ is a solution to the minimization problem~\eqref{variationnal_problem_squarred_grad}. $\lambda$ is a real number and $\delta f$ is a function such that $\delta f(\textbf{q}) = 0$ for any $\textbf{q}\in \overline{R} \cup \overline{P}$. Since for any $\lambda$, 
$$\int_{\left(\overline{R} \cup \overline{P}\right)^\mathrm{c}} |\nabla f_\lambda|^2 \exp(-\beta V) \ge
\int_{\left(\overline{R} \cup \overline{P}\right)^\mathrm{c}} |\nabla f^*|^2 \exp(-\beta V),$$ 
the solution verifies:
$$\frac{\mathrm{d}}{\mathrm{d}\lambda} \left(\int_{\left(\overline{R} \cup \overline{P}\right)^\mathrm{c}} |\nabla f_\lambda|^2 \exp(-\beta V) \right)= 0.$$ 
Thus,
$f^*$ satisfies the following Euler equations:
\begin{equation}
    \label{squarred_grad_proof_1}
    \begin{aligned}
        0 = & \left. \frac{1}{2} \frac{\mathrm{d}}{\mathrm{d} \lambda} \left(\int_{\left(\overline{R} \cup \overline{P}\right)^\mathrm{c}} \left|\nabla f_\lambda(\textbf{q})  \right|^2 \mathrm{e}^{-\beta V(\textbf{q})} \, \mathrm{d}\textbf{q} \right)\right|_{\lambda=0} \\
        = &  \left. \frac{1}{2} \frac{\mathrm{d}}{\mathrm{d} \lambda} \left(\int_{\left(\overline{R} \cup \overline{P}\right)^\mathrm{c}}\left( \left|\nabla f^*(\textbf{q})\right|^2 + 2\lambda \nabla \delta f(\textbf{q}) \cdot \nabla f^*(\textbf{q}) + \lambda^2 \left|\nabla \delta f(\textbf{q})\right|^2\right) \mathrm{e}^{-\beta V(\textbf{q})} \,\mathrm{d}\textbf{q} \right)\right|_{\lambda=0} \\
        = & \int_{\left(\overline{R} \cup \overline{P}\right)^\mathrm{c}} \nabla \delta f(\textbf{q}) \cdot \nabla f^*(\textbf{q}) \mathrm{e}^{-\beta V(\textbf{q})} \, \mathrm{d}\textbf{q}.
    \end{aligned}
\end{equation}
Therefore, 
\begin{equation*}
    \int_{\left(\overline{R} \cup \overline{P}\right)^\mathrm{c}} \nabla \cdot \left( \delta f(\textbf{q}) \nabla f^*(\textbf{q}) \mathrm{e}^{-\beta V(\textbf{q})} \right)\, \mathrm{d}\textbf{q} = \int_{\left(\overline{R} \cup \overline{P}\right)^\mathrm{c}} \delta f(\textbf{q}) \nabla \cdot \left(  \nabla f^*(\textbf{q}) \mathrm{e}^{-\beta V(\textbf{q})} \right)\,\mathrm{d}\textbf{q}.
\end{equation*}
Now, the term on the left hand side can be rewritten as a surface integral:
\begin{equation*}
    \label{condtion_IPP_simplification}
    \int_{\left(\overline{R} \cup \overline{P}\right)^\mathrm{c}} \nabla \cdot \left( \delta f(\textbf{q}) \nabla f^*(\textbf{q}) \mathrm{e}^{-\beta V(\textbf{q})} \right) \, \mathrm{d}\textbf{q} = \int_{\partial \left(\left(\overline{R} \cup \overline{P}\right)^\mathrm{c} \right) } \delta f(\textbf{q}) \nabla f^*(\textbf{q}) \cdot \textbf{n} \, \mathrm{e}^{-\beta V(\textbf{q})} \,\mathrm{d}s = 0,
\end{equation*}
where $\textbf{n}$ is the outgoing unit normal vector to $\left(\overline{R} \cup \overline{P}\right)^\mathrm{c}$. This integral vanishes since $\delta f=0$ on $\partial R\cup\partial P$.  Finally, we obtain:
\begin{equation}
    \label{squarred_grad_proof_2}
    \begin{aligned}
        0 = & -  \int_{\left(\overline{R} \cup \overline{P}\right)^\mathrm{c}} \delta f(\textbf{q}) \nabla \cdot \left(  \nabla f^*(\textbf{q}) \, \mathrm{e}^{-\beta V(\textbf{q})} \right) \, \mathrm{d}\textbf{q} \\
         = & -  \beta \int_{\left(\overline{R} \cup \overline{P}\right)^\mathrm{c}} \delta f(\textbf{q})  (\mathcal{L}_\mathrm{ovd} f^*)(\textbf{q}) \, \mathrm{e}^{-\beta V(\textbf{q})} \,\mathrm{d}\textbf{q}.
    \end{aligned}
\end{equation}
Since the latter integral vanishes for any function $\delta f$, we can conclude that $\mathcal{L}_\mathrm{ovd} f^* = 0$, and so $f^*$ is the committor function since $f^*$ satisfies the Dirichlet boundary conditions (see equation~\eqref{back_kolmo_committor} above). 

\section{Proof that the committor is the only solution to equation~(20)}
\label{sec:fixed_point_proof}

We first recall here equation~(20) of the main text: 
\begin{equation}
    \label{committor_fixde_point_2}
    \forall t > 0, \forall \textbf{q} \in \left(\overline{R} \cup \overline{P}\right)^\mathrm{c}, \quad \left( \mathrm{Id} - \mathcal{P}_t^\mathrm{i} \right) \chi (\textbf{q}) - \left( \mathcal{P}_t^\mathrm{b} \mathbb{1}_P \right)(\textbf{q}) = 0.
\end{equation}
Let us consider two solutions $p_1$ and $p_2$ of~\eqref{committor_fixde_point_2}. Then, 
\begin{equation*}
    p_1 - p_2 = \mathcal{P}_t^\mathrm{i} \left(p_1 - p_2 \right).
\end{equation*}
By applying the operator $\mathcal{P}_t^\mathrm{i}$ on $(p_1 - p_2)$ $n$ times, we then obtain: for any $\textbf{q}_0 \in \left(\overline{R} \cup \overline{P} \right)^c$
\begin{equation*}
    \forall n \in \mathbb{N}, \, (p_1 - p_2)(\textbf{q}_0) = \left(\mathcal{P}_t^\mathrm{i}\right)^n \left(p_1 - p_2 \right)(\textbf{q}_0) = \mathcal{P}_{nt}^\mathrm{i}\left(p_1 - p_2 \right)(\textbf{q}_0) = \mathbb{E}^{\textbf{q}_0}\left[\left(p_1 - p_2\right)(\textbf{q}_{nt})\mathbb{1}_{nt < \tau}\right],
\end{equation*}
thanks to the semigroup property $\mathcal{P}_t^\mathrm{i}  \mathcal{P}_s^\mathrm{i} = \mathcal{P}_{t+s}^\mathrm{i}$.
In particular,
\[
|p_1(\textbf{q}_0) - p_2(\textbf{q}_0)| \leq 2 \mathbb{E}^{\textbf{q}_0}[\mathbb{1}_{nt < \tau}].
\]
Finally, considering the limit $n \to \infty$, as $\tau$ is almost surely finite, we have $\underset{n\rightarrow +\infty}{\lim} \mathbb{1}_{nt < \tau} = 0$ almost surely and thus we obtain $p_1 = p_2$ by dominated convergence.

\section{Derivation of the discrete time committor fixed point equation}
\label{sec:discrete_committor}

Let us introduce the transition operator of the original continuous in time process (see Eq.~(21) of the main text). The trajectory is denoted by $(\textbf{q}_t)_{t \geqslant 0}$ and for any bounded measurable function $f : \Omega \to \mathbb{R}$,
\[
\left(\mathcal {Q}_{t} f \right) (\textbf{q}) = \mathbb{E}\left[f\left(\textbf{q}_{t}\right) \middle| \textbf{q}_0 = \textbf{q}\right].
\]
Notice that, for any $t > 0$, $\mathcal {Q}_{t}$ is symmetric with respect to $\mathrm{e}^{-\beta V(\textbf{q})} \, \mathrm{d} \textbf{q}$.

Let us now introduce a fixed time-lag $\vartheta >0$. We consider the committor function associated with the Markov Chain $(\textbf{q}_n^{\vartheta})_{n \ge 0}$, with transition kernel $\mathcal{Q}_{\vartheta}$. In practice, this transition kernel is approximated by $L \geqslant 1$ steps of the Markov chain with time step $\Delta t = \frac{\vartheta}{L}$: $\mathcal{Q}_{\vartheta} \varphi(\textbf{q}) \approx \mathbb{E} \left[\varphi(Q^\ell) \middle| Q^0 = \textbf{q} \right]$, where $\left(Q^\ell\right)_{0 \leqslant \ell \leqslant L}$ satisfies
\[
\forall \ell \in \{0, \cdots, L - 1 \}, \quad  Q^{\ell+1} = Q^\ell - \nabla V (Q^\ell) \Delta t + \sqrt{\frac{2\Delta t}{\beta}}G^\ell,
\]
It would be interesting to analyze the error introduced by this approximation, but this is beyond the scope of this work.

The first hitting times of $R$ and $P$ for the Markov chain $(Q_{n\vartheta})_{n \ge 0}$ are defined as
\begin{equation*}
T_R  = {\inf}\left\{n \geqslant 0 \, \middle|  \, \textbf{q}_n^{\vartheta} \in R \right\}, \text{ and }
T_P  = {\inf}\left\{n \geqslant 0 \, \middle| \, \textbf{q}_n^{\vartheta} \in P \right\}.
\end{equation*}
The associated committor function is
\begin{equation}\label{eq:def_discrete_comm}
\forall \textbf{q} \in \Omega, \qquad  \chi_\vartheta(\textbf{q})  = \mathbb{P}\left(T_P < T_R \, \middle| \, \textbf{q}_0 = \textbf{q}\right).
\end{equation}
Notice that, by definition,
\begin{equation}
\label{eq:boundary_cdt_discrete_comm}
\chi_\vartheta(\textbf{q}) =
\left\{
\begin{aligned}
    0 & \text{ if } \textbf{q} \in R,\\
    1 & \text{ if } \textbf{q} \in P.
\end{aligned}
\right.
\end{equation}
The function $\chi_\vartheta$ is smooth on $\Omega$ except on $\partial R \cup \partial P$ where it is discontinuous.

Let us now check that $\chi_\vartheta$ is solution to a fixed point equation. For $\textbf{q} \in \left(\overline{R} \cup \overline{P}\right)^\mathrm{c}$,
\begin{align*}
    \chi_\vartheta(\textbf{q}) &=\mathbb{E}\left(\mathbb{1}_{\{T_P < T_R\}} \, \middle| \, \textbf{q}_0^\vartheta = \textbf{q}\right)\\
    &=\mathbb{E}\left(\mathbb{1}_{\{\textbf{q}_1^\vartheta \in R\}}\mathbb{1}_{\{T_P < T_R\}} \,\middle| \, \textbf{q}_0^\vartheta = \textbf{q}\right)+\mathbb{E}\left(\mathbb{1}_{\{\textbf{q}_1^\vartheta \in P\}}\mathbb{1}_{\{T_P < T_R\}} \, \middle| \, \textbf{q}_0^\vartheta = \textbf{q}\right)\\
    & \quad +\mathbb{E}\left(\mathbb{1}_{\{\textbf{q}_1^\vartheta \in \left(\overline{R} \cup \overline{P}\right)^\mathrm{c}\}}\mathbb{1}_{\{T_P < T_R\}} \, \middle| \, \textbf{q}_0^\vartheta = \textbf{q}\right)\\
    &= 0 + \mathbb{P}\left(\textbf{q}_1^\vartheta \in P \, \middle| \, \textbf{q}_0^\vartheta = \textbf{q}\right) + \mathbb{E} \left( \chi_\vartheta(\textbf{q}_1^\vartheta) 1_{\{\textbf{q}_1^\vartheta \in \left(\overline{R} \cup \overline{P}\right)^\mathrm{c}\}} \, \middle| \, \textbf{q}_0^\vartheta = \textbf{q}\right) \\ 
    &= \mathbb{E} \left( \chi_{\vartheta}(\textbf{q}_1^\vartheta) \middle| \textbf{q}_0^\vartheta =\textbf{q} \right),
\end{align*}
where we used a conditioning by $Q_{\vartheta}$ in the last line. We thus obtain:
\begin{equation}\label{eq:fp_discrete_comm}
\forall \textbf{q} \in \left(\overline{R} \cup \overline{P}\right)^\mathrm{c}, \, \chi_\vartheta(\textbf{q}) = \mathcal Q_{\vartheta} (\chi_\vartheta)(\textbf{q}).
\end{equation}
Notice that $\chi_\vartheta$ is actually uniquely defined by~\eqref{eq:boundary_cdt_discrete_comm}--\eqref{eq:fp_discrete_comm}, since this is equivalent to~\eqref{eq:def_discrete_comm}. 

\section{Proof that the committor solves the minimization problem~(25)}
\label{sec:proof_variational_transition_operator}

Recall the minimization problem~(25) of the main text: 
\begin{equation}
    \label{minimization_fixed_point_canonical}
        \underset{f}{\mathrm{arginf}} \left\{ \frac{1}{2} \int_{\left(\overline{R} \cup \overline{P}\right)^\mathrm{c}}  f(\textbf{q})\left(\mathrm{Id} - \mathcal{P}_t^\mathrm{i} \right) f (\textbf{q}) \, \mathrm{e}^{-\beta V(\textbf{q})} \, \mathrm{d}\textbf{q}  - \int_{\left(\overline{R} \cup \overline{P}\right)^\mathrm{c}} f(\textbf{q}) \mathcal{P}_t^\mathrm{b} \mathbb{1}_{P}(\textbf{q}) \, \mathrm{e}^{-\beta V(\textbf{q})} \, \mathrm{d}\textbf{q} \right\}.
\end{equation}
Let us prove following the approach presented in Ref.~\onlinecite{Li2021} that the committor function solves this minimization problem. Let $f^*$ be a minimizing function solution to~\eqref{minimization_fixed_point_canonical}. As in Section 1 above, let us write the Euler equations satisfied by $f^*$.By considering $f_\lambda(\textbf{q})~=~f^*(\textbf{q})~+~\lambda\delta f(\textbf{q})$ for an arbitrary smooth function $\delta f$, 
\begin{equation}
    \label{fixed_point_canonical_proof}
    \begin{aligned}
        0 = & \, \frac{\mathrm{d}}{\mathrm{d} \lambda} \Bigg( \frac{1}{2} \int_{\left(\overline{R} \cup \overline{P}\right)^\mathrm{c}} f_\lambda (\textbf{q}) (\mathrm{Id} - \mathcal{P}_t^\mathrm{i} ) f_\lambda (\textbf{q}) \, \mathrm{e}^{-\beta V(\textbf{q})} \, \mathrm{d}\textbf{q}  \;   \\ & \qquad \qquad \qquad \qquad - \int_{\left(\overline{R} \cup \overline{P}\right)^\mathrm{c}}f_\lambda (\textbf{q}) \mathcal{P}_t^\mathrm{b} \mathbb{1}_{P} (\textbf{q}) \, \mathrm{e}^{-\beta V(\textbf{q})} \, \mathrm{d}\textbf{q} \Bigg) \Bigg|_{\lambda=0} \\
        = & \int_{\left(\overline{R} \cup \overline{P}\right)^\mathrm{c}} \delta f(\textbf{q})\left(\mathrm{Id} - \mathcal{P}_t^\mathrm{i} \right) f^*(\textbf{q}) \, \mathrm{e}^{-\beta V(\textbf{q})}\mathrm{d}\textbf{q}  -  \int_{\left(\overline{R} \cup \overline{P}\right)^\mathrm{c}} \delta f(\textbf{q}) \mathcal{P}_t^\mathrm{b} \mathbb{1}_{P} (\textbf{q})\, \mathrm{e}^{-\beta V(\textbf{q})} \, \mathrm{d}\textbf{q} \\
        = & \int_{\left(\overline{R} \cup \overline{P}\right)^\mathrm{c}} \delta f(\textbf{q}) \left[\left(\mathrm{Id} - \mathcal{P}_t^\mathrm{i} \right) f^*(\textbf{q}) -  \mathcal{P}_t^\mathrm{b} \mathbb{1}_{P} (\textbf{q})\right] \, \mathrm{e}^{-\beta V(\textbf{q})} \, \mathrm{d}\textbf{q}.
    \end{aligned}
\end{equation}
The second equality is justified by the fact that the operator $\mathcal{P}_t^\mathrm{i}$ is symmetric with respect to $\mathrm{e}^{-\beta V(\textbf{q})} \, \mathrm{d}\textbf{q}$ restricted to $\left(\overline{R} \cup \overline{P}\right)^\mathrm{c}$, i.e. for any functions $u$ and $v$ from $\left(\overline{R} \cup \overline{P}\right)^\mathrm{c}$~to~$\mathbb{R}$, it holds
\begin{equation}
    \label{internal_propagator_simmetry}
    \int_{\left(\overline{R} \cup \overline{P}\right)^\mathrm{c}} u(\textbf{q}) \left(\mathcal{P}_t^\mathrm{i} v\right) (\textbf{q}) \, \mathrm{e}^{-\beta V(\textbf{q})} \, \mathrm{d}\textbf{q} = \int_{\left(\overline{R} \cup \overline{P}\right)^\mathrm{c}} v(\textbf{q}) \left(\mathcal{P}_t^\mathrm{i} u\right) (\textbf{q}) \, \mathrm{e}^{-\beta V(\textbf{q})} \, \mathrm{d}\textbf{q}.
\end{equation}
A proof of this symmetry property can be found in Ref. \onlinecite{Li2021}. As $\delta f$ is an arbitrary function, \eqref{fixed_point_canonical_proof} implies that 
$$
\forall \textbf{q} \in \left(\overline{R} \cup \overline{P}\right)^\mathrm{c}, \left(\mathrm{Id} - \mathcal{P}_t^\mathrm{i} \right) f^*(\textbf{q}) -  \mathcal{P}_t^\mathrm{b} \mathbb{1}_{P} (\textbf{q}) = 0,
$$
and the only solution to this equation is the committor function as justified in~\ref{sec:fixed_point_proof}.

\section{Proof that the discrete committor function solves the minimization problem~(28)}
\label{sec:proof_committor_roux}

Let us now show that $\chi_\vartheta$ is the solution to the minimization problem~(28) of the main text which we recall here:
\begin{equation}
 \label{eq:minimization_fixed_point_roux}
    \underset{f}{\mathrm{inf}} \left\{\int_{\Omega} \left( \mathcal {Q}_{\vartheta} \left(f^2\right)(\textbf{q})  -2 f(\textbf{q})\mathcal {Q}_{\vartheta} f(\textbf{q}) +f(\textbf{q})^2\right) \mathrm{e}^{-\beta V(\textbf{q})} \, \mathrm{d}\textbf{q}  \, \middle| \, f(\textbf{q}) = \mathbb{1}_P(\textbf{q}), \, \forall \textbf{q} \in \left(\overline{R} \cup \overline{P}\right)  \right\}.
\end{equation}
Notice that the infimum is well defined as the infimum of a non negative quantity. Indeed,
$$
\int_{\Omega} \left( \mathcal {Q}_{\vartheta} \left(f^2\right)(\textbf{q})  -2 f(\textbf{q})\mathcal {Q}_{\vartheta} f(\textbf{q}) +f(\textbf{q})^2\right) \mathrm{e}^{-\beta V(\textbf{q})} \, \mathrm{d}\textbf{q} = \mathbb E\left[ (f(\textbf{q}_{\vartheta}) - f(\textbf{q}_0))^2 \right]
$$
where f$(\textbf{q}_t)_{t \ge 0}$ is a solution to the overdamped Langevin dynamics (see Eq. (3) of the main text) starting at equilibrium: $\textbf{q}_0 \sim \mathrm{e}^{-\beta V(\textbf{q})} \, \mathrm{d}\textbf{q}$.
Since $\mathcal Q_{\vartheta}$ preserves the measure $\mathrm{e}^{-\beta V(\textbf{q})/Z}$, we have, for any function $f$,
\begin{equation}
    \int_{\Omega} \left( \mathcal{Q}_{\vartheta} \left(f^2\right) (\textbf{q})  -2 f(\textbf{q})\mathcal{Q}_{\vartheta} f(\textbf{q}) +f(\textbf{q})^2\right) \mathrm{e}^{-\beta V(\textbf{q})} \, \mathrm{d}\textbf{q}  = 2 \int_{\Omega} f(\textbf{q})\left(\mathrm{Id}  - \mathcal{Q}_{\vartheta} \right)f(\textbf{q}) \, \mathrm{e}^{-\beta V(\textbf{q})} \, \mathrm{d}\textbf{q}.
\end{equation} 
Let us now write  the Euler equation satisfied by a solution $f^*$ to the minimization problem~\eqref{eq:minimization_fixed_point_roux}. By considering $f_\lambda(\textbf{q})~=~f^*(\textbf{q})~+~\lambda\delta f(\textbf{q})$ for an arbitrary smooth function $\delta f$ vanishing on $\overline{R} \cup \overline{P}$, we can write,  using the fact that $\mathcal{Q}_{\vartheta}$ is reversible with respect to $Z^{-1}\mathrm{e}^{-\beta V(\textbf{q})} \, \mathrm{d}\textbf{q}$: 
\begin{equation}
\begin{aligned}
    0 &= \left.\frac{\mathrm{d}}{\mathrm{d} \lambda} \left(\int_{\left(\overline{R} \cup \overline{P}\right)^\mathrm{c}} f_\lambda(\textbf{q})\left(\mathrm{Id}  - \mathcal{Q}_{\vartheta} \right)f_\lambda(\textbf{q}) \, \mathrm{e}^{-\beta V(\textbf{q})} \, \mathrm{d}\textbf{q}\right)\right|_{\lambda=0} \\ 
    & = 2 \int_{\left(\overline{R} \cup \overline{P}\right)^\mathrm{c}} \delta f (\textbf{q})\left(\mathrm{Id}  - \mathcal{Q}_{\vartheta} \right)f^*(\textbf{q}) \, \mathrm{e}^{-\beta V(\textbf{q})} \, \mathrm{d}\textbf{q}.
\end{aligned}
\end{equation}
Since $\delta f$ is arbitrary, one gets $(\mathrm{Id}  - \mathcal{Q}_{\vartheta}) f^*(\textbf{q})=0$  for $\textbf{q} \in \left(\overline{R} \cup \overline{P}\right)^\mathrm{c}$
and thus, since $f^*$ satisfies the boundary conditions~\eqref{eq:boundary_cdt_discrete_comm}, one finally gets $f^*=\chi_\vartheta$.

\section{Adaptive multilevel splitting for molecular dynamics}
\label{sec:AMS}

Adaptive multilevel splitting (AMS) is an algorithm designed construct automatically a splitting estimator, adequate to estimate rare event probabilities. In the case of sampling rare reactive trajectories starting in $R$ and ending in $P$, to build a splitting estimator, one needs to introduce a sequence of interfaces
\[
\Sigma_0 = \partial R, \, \Sigma_1, \, \dots, \, \Sigma_M = \partial P,
\]
such that any reactive trajectory must cross all interfaces in order. The estimator of the committor probability averaged on the boundary can then be written as:
\begin{equation}
\label{eq:prob_product}
\chi(\partial R) \;=\; p_{R\to \Sigma_1}(\partial R) \,\prod_{j=1}^{M-1} p_{R \to \Sigma_{j+1}}(\Sigma_j)\; \times\; p_{R \to P}(\Sigma_M),
\end{equation}
where $p_{R \to \Sigma_{j+1}}(\Sigma_j)$ denotes the probability to hit $\Sigma_{j+1}$ before returning to $R$, conditional on first hitting $\Sigma_j$.

AMS is a multiple replica algorithm. It starts with $N_{\mathrm{rep}}$ replicas starting on the exit set of $\partial R$, propagated until each replica hits either $R$ or $P$. When considering the overdamped dynamics, to define the exit set of $R$, on need in fact to introduce a surface $\Sigma_R$ enclosing the state $A$ such that a trajectory exiting $R$ first cross $\partial R$ and then $\Sigma_R$. For each replica, a progress score is defined as the maximum value $z_\mathrm{max}$ of a one dimensional reaction coordinate $\xi$ along the trajectory. At iteration $j$, replicas are first sorted by increasing $z_\mathrm{max}$, then, the algorithm kills the fraction $\eta_j^{\mathrm{killed}}/N_{\mathrm{rep}}$ with the $k^\mathrm{th}$ lowest score $z_\mathrm{kill} = z^k_\mathrm{max}$ and replace them by clones of one of the remaining ones branched at $z_\mathrm{kill}$. The dynamics is then propagated until the system reaches $R$ or $P$. Each iteration is equivalent to placing a surface $\Sigma_j$ such that the probability $ p_{R \to \Sigma_{j}}(\Sigma_{j-1}) = 1-\frac{\eta_j^{\mathrm{killed}}}{N_{\mathrm{rep}}}$ The splitting estimator of the boundary-averaged committor is
\begin{equation}
\label{eq:AMS_proba_estimator}
\widehat{p}_{\Sigma_j\to\Sigma_{j+1}}\left( \Sigma_j \right)
= 1-\frac{\eta_j^{\mathrm{killed}}}{N_{\mathrm{rep}}},
\qquad
\widehat{p}^{\mathrm{AMS}}
=\prod_{j=1}^{n_{\max}}\!\left(1-\frac{\eta_j^{\mathrm{killed}}}{N_{\mathrm{rep}}}\right),
\end{equation}
where $n_{\max}$ is the number of iterations of AMS algorithm done until all the replicas end in $P$. AMS places the “interfaces” adaptively, and assuming that the fraction with the worst score is constant, meaning that there are always $k$ replicas killed at each iteration, the conditional probabilities are thus equalized, which reduces the variance of the estimator.

\bibliography{committor}